\documentclass[fleqn,usenatbib]{mnras}

\usepackage{newtxtext,newtxmath}
\usepackage{gensymb}
\usepackage[T1]{fontenc}
\usepackage{ae,aecompl}

\usepackage{graphicx}	% Including figure files
\usepackage{amsmath}	% Advanced maths commands
\usepackage{longtable}
\usepackage{booktabs}
\usepackage{tabularx}
\usepackage{supertabular}

\title[Spin--filament alignments at $z\sim0.3$]{The MAGPI Survey: the evolution of galaxy spin--filament alignments since $\mathbf{z}$$\sim$0.3 }

\author[S.\ Barsanti et al.]
{Stefania Barsanti$^{1}$\thanks{E-mail: stefania.barsanti@sydney.edu.au},
Andrew J. Battisti$^{2,3}$,
Sabine Bellstedt$^{2}$,
Matthew Colless$^{3}$,
\newauthor
Scott M. Croom$^{1}$,
Caroline Foster$^{4}$,
Katherine E. Harborne$^{5,6}$,
Claudia D. P. Lagos$^{2}$,
\newauthor
J. Trevor Mendel$^{3}$,
Matt S. Owers$^{7,8}$,
Elmo Tempel$^{9}$,
Emily Wisnioski$^{3}$,
Iris Breda$^{10}$,
\newauthor
Pratyush K. Das$^{11}$,
Clotilde Laigle$^{12}$,
Rhea-Silvia Remus$^{13,14}$,
Gauri Sharma$^{15,16,17,18,19}$,
\newauthor
Sabine Thater$^{20}$,
Lucas M. Valenzuela$^{13}$,
Jesse van de Sande$^{4}$,
Glenn van de Ven$^{20}$,
\newauthor
Daria Zakharova$^{21}$,
Bodo L. Ziegler$^{20}$
\\ 
\\ 
Affiliations are listed at the end of the paper
}

\date{Accepted 2026 October 28. Received 2026 October 25; in original form 2026 September 10}

\pubyear{2026}

\begin{document}
\label{firstpage}
\pagerange{\pageref{firstpage}--\pageref{lastpage}}
\maketitle

\begin{abstract}
We present measurements of galaxy spin–filament alignments at 
0.20\,$<$\,$z$\,$<$\,0.42 using the MAGPI Survey, exploiting resolved stellar and ionised gas kinematics to reconstruct 3D spin axes. Cosmic filaments are identified using a stellar mass-matched constant-number-density population from the GAMA, MAGPI Redshift Survey and DESI datasets. We detect a flipping of the statistical alignment signal at $z$\,$\sim$\,0.3 as a function of $M_{\star}$: low-mass galaxies preferentially exhibit spins parallel to the direction of their host filaments, while high-mass galaxies tend towards perpendicular orientations. The transition mass is measured to be $\log(M_{\star,t}/M_{\odot})$\,$\simeq$\,10.8, approximately 0.6\,dex higher than the value for the SAMI Galaxy Survey at $z$\,$\sim$\,0.05. Bootstrap analysis gives 99.5\% confidence that the transition mass is higher for MAGPI than SAMI. At fixed stellar mass MAGPI galaxies exhibit stronger parallel alignments than their SAMI counterparts, suggesting evolution of spin--filament alignments over the past $\sim$\,3\,Gyr. The gaseous component in MAGPI galaxies is more closely aligned with filaments than the stellar component, relative to a $M_{\star}$-matched SAMI sample. This indicates a stronger coupling between ionised gas and the large-scale structure at $z$\,$\sim$\,0.3, consistent with coherent gas accretion onto filaments remaining efficient at higher $M_{\star}$. At $z$\,$\sim$\,0 we detect a transition toward perpendicular alignments at lower $M_{\star}$, while the gas and stellar components exhibit similar alignments relative to filaments, likely due to the cumulative effects of mergers and secular evolution.
\end{abstract}

%It should be a single paragraph not more than 250 words (200 words for Letters).
%No references should appear in the abstract.


%%%%%%%%%%%%%%%%% BODY OF PAPER %%%%%%%%%%%%%%%%%%

\section{Introduction}
\label{Introduction}

Understanding how galaxies acquire their angular momentum is fundamental for a complete picture of galaxy formation and evolution. According to tidal torque theory, galaxies acquire their spin through torques exerted by the surrounding large-scale structure during the early phases of halo collapse, leading to correlations between galaxy angular momentum and the cosmic web  \citep{Hoyle1951,Peebles1969,Doroshkevich1970,White1984,Porciani2002,Schafer2009}. Detecting these correlations can help us to provide insights into the processes driving galaxy growth.

Large-scale cosmological hydrodynamic simulations find that the orientation between the galaxy spins and their closest filament depends on stellar mass \citep{Dubois2014,Laigle2015,Codis2018,Wang2018,Kraljic2020}. Low-mass galaxies tend to align parallel to direction of their closest filament, consistent with growth dominated by smooth gas accretion, whereas more massive galaxies show perpendicular alignments associated with merger-driven angular momentum reorientation \citep{Welker2014}. The transition between parallel and perpendicular alignments is therefore linked to a `transition mass' that has been found to be redshift-dependent for dark matter halos \citep{Codis2012,Codis2015,Tobar2026}. Different galaxy formation models predict different transition masses between parallel and perpendicular orientations \citep{Krolewski2019,Kraljic2019,GaneshaiahVeena2019}, but this transition mass so far has remained unconstrained beyond $z$\,$\sim$\,0 by current spin observations.

At $z$\,$\sim$\,0 the transition mass for galaxy spin--filament alignments has been explored observationally through spatially-resolved stellar kinematics \citep{Welker2020,Barsanti2022} and via HI studies \citep{BlueBird2020,Tudorache2022}. Taking advantage of the SAMI Galaxy Survey, \citet{Welker2020} measured the transition stellar mass in the range $\log(M_\star/M_{\odot})=10.4-10.9$ for the first time. Subsequently, \citet{Barsanti2022,Barsanti2023} extended this analysis by investigating stellar and ionised-gas spin--filament alignments separately and detected the flipping of the statistical alignment signal already at $\log(M_\star/M_{\odot})=10.0$. 

Results from the Horizon-AGN simulation predict a redshift evolution of spin–filament alignments \citep{Codis2018}: the alignment of low-mass galaxies diminishes towards low redshifts, while the perpendicular spin orientation of massive galaxies strengthens with time, likely driven by mergers. At fixed mass, these alignments also depend on galaxy morphology, with high-redshift signals dominated by disk galaxies, whereas elliptical centrals primarily contribute to the perpendicular alignment. \cite{Barsanti2022} showed that the property of individual galaxies with which spin alignment correlates most strongly is bulge mass: galaxies having higher bulge masses tend to have spins perpendicular to their host filaments. 

Cosmological N-body simulations have predicted the orientation of the dark matter halo spin vector to be mass-dependent (\citealp{AragonCalvo2007,Codis2012,Trowland2013,GaneshaiahVeena2018}). However, recently, \citet{Tobar2026} found no redshift evolution in dark matter halo spin–filament alignments in  dark matter-only zoom-in N-body simulations covering $0 < z < 1$, suggesting that the present-day spin configuration is largely established at earlier stages of halo assembly. Observationally, extending such measurements beyond $z$\,$\sim$\,0 remains challenging. Recent studies of galaxy shape–filament alignments have been enabled by Euclid at $0.5 < z < 0.9$ \citep{Laigle2025}. Using the Euclid Quick Data Release 1 and reconstructing the cosmic web from photometric redshifts, they found that massive early-type galaxies ($\log(M_\star/M_{\odot}) > 10$) exhibit a preferential alignment of their major axes along projected cosmic filaments, in agreement with the low-redshift high-mass galaxies.

Galaxy spin–filament alignments provide a powerful tracer of galaxy assembly history, and extending these studies beyond $z$\,$\sim$\,0 offers a critical observational test of galaxy evolution models. Measuring this evolution is critical, as it probes how galaxies acquire, redistribute, and reorient their angular momentum across cosmic time and as a function of mass. 

In this work, we extend spin-filament alignment studies beyond $z$\,$\sim$\,0 for the first time using spatially-resolved kinematics. We explore the transition stellar mass for galaxy spin--filament alignments at $z\sim0.3$, taking advantage of the Middle Ages Galaxy Properties with Integral field spectroscopy (MAGPI; \citealp{Foster2021}) survey. MAGPI is a Large Program on the European Southern Observatory Very Large Telescope using the MUSE instrument \citep{Bacon2010}, designed to study the physical drivers of galaxy transformation at a lookback time of 3–4 Gyr ($0.26\leq z \leq 0.42$). By comparing MAGPI galaxy spin--filament alignments with the results from the SAMI Galaxy Survey, we investigate the potential redshift evolution of these alignments and their transition stellar mass. 

This paper is organised as follows. We present our MAGPI galaxy sample, including stellar and ionised gas kinematics and galaxy parameters in Section~\ref{Data and Galaxy Sample}. We also provide a brief summary of the SAMI Galaxy Survey and the $z$\,$\sim$\,0 reference sample for galaxy spin–filament alignments from \citet{Barsanti2022}. In Section~\ref{Reconstruction of the cosmic web} we describe the reconstruction of the cosmic filaments. In Section~\ref{Methods} we identify the orientation of the spin axes. In Section~\ref{Results} we present our results on the alignments, their correlations with stellar mass, the measured transition mass, and the comparison with SAMI galaxy spin--filament alignments at $z$\,$\sim$\,0. In Section~\ref{Discussion} we compare our findings to previous studies, and we discuss their physical interpretations. Finally, we summarise our findings and state our conclusions in Section~\ref{Summary and conclusions}. Throughout this work, we assume $\Omega_{m}=0.3$, $\Omega_{\Lambda}=0.7$ and $H_{0}=70$\,km\,s$^{-1}$\,Mpc$^{-1}$ as the cosmological parameters.

\section{Data and Galaxy Sample}
\label{Data and Galaxy Sample}

\subsection{The MAGPI Survey}
\label{The MAGPI Survey}

The Middle Ages Galaxy Properties with Integral field spectroscopy (MAGPI) survey is a VLT/MUSE Large Program of 340~hr \citep[][Program ID: 1104.B-0536]{Foster2021, Mendel2024}. The survey focusses on galaxies at a look-back time of 3--4\,Gyr ($z\sim0.3$), aiming to understand the evolution of the interplay between internal and external processes shaping galaxy properties. MAGPI uses ground-layer adaptive optics (GLAO), which provides excellent spatial resolution ($i$-band PSF FWHM$\sim$0.55\arcsec). It exploits the wide-field mode of MUSE, which has a 1 $\times$ 1 arcmin$^{2}$ field-of-view (FOV) sampled at 0.2 $\times$ 0.2 arcsec$^{2}$ and covers a wavelength range between 4700 and 9351 \AA\ with a spectral resolution of $R = 2728$ at 7025 \AA.

The MAGPI Survey targets 60 massive central galaxies (primary galaxies) and their neighbours within the MUSE FOV, chosen to cover uniformly different environments (field, groups and clusters) at $0.26\leq z\leq 0.42$. For this redshift range all strong rest-frame optical emission and absorption lines are covered by the MUSE instrument to probe ionised gas and stellar properties \citep{Battisti2026}. A total of 56 primary galaxies were selected from the Galaxy And Mass Assembly survey \citep[GAMA;][]{Driver2011} G12, G15 and G23 regions, defining 56 independent fields (56 arcmin$^2$). They cover a range of halo mass $11.35 \leq \textup{log} (\textup{M}_\textup{halo}/\textup{M}_{\odot})\leq 15.35$ \citep{Robotham2011}. The remaining 4 primary galaxies were chosen from MUSE archival observations of Abell 370 and Abell 2744 clusters. The survey has completed the observations of all 56 fields, which have been fully reduced with relevant data products. In this work, we focus on the large-scale structure surrounding the 56 primary galaxies, excluding the two clusters. 

MAGPI MUSE data reduction is summarised in \cite{Foster2021} and \cite{Foster2025}, and the data release will be presented by Mendel et al. (in prep). Briefly, raw MUSE datacubes are reduced using the \texttt{PYMUSEPIPE}\footnote{\url{https://github.com/emsellem/pymusepipe}} software \citep{Emsellem2022}, a Python wrapper for the MUSE reduction pipeline. The pipeline is used to perform the standard bias and overscan subtraction, flat-fielding, wavelength calibration, telluric correction and cube reconstruction steps. The Zurich Atmosphere Purge (ZAP; \citealp{Soto2016}) package is used to improve background sky subtraction. Since the MUSE cubes are the deepest images available and to ensure uniformity in detecting targets, galaxies and other objects are detected directly on the white light MUSE image using the image analysis and source extraction package {\sc ProFound} \citep{Robotham2018}. Segmentation maps are created with {\sc ProFound} to define the edges of every detected source. The segmentation maps are used to cut the main MUSE datacube into minicubes for every detected object, produced using the software package \texttt{MPDAF}\footnote{\url{https://github.com/musevlt/mpdaf}}. For each identified object, a spectroscopic redshift is measured using the \texttt{MARZ} software\footnote{\url{https://github.com/Samreay/Marz}} \citep{Hinton2016}, using both an initial automated estimate and subsequent visual inspection. 

{\sc ProFound} also provides magnitudes in $g$, $r$ and $i$ in synthetic Sloan Digital Sky Survey filters. Galaxy structural parameters, such as effective radius and axial ratio, in all three synthetic bands are obtained using both {\sc ProFound} and the photometric fitting parametric light profiles tool {\sc GalFit} \citep{Peng2002, Peng2010}. 

\subsection{Stellar kinematics}
\label{Stellar kinematics}

To identify the stellar spin axis, we take advantage of spatially-resolved stellar kinematics. A description of the stellar kinematic MAGPI data products is presented in \cite{DEugenio2023}, \cite{Derkenne2024} and \cite{Foster2025}. In summary, stellar mean velocity and velocity dispersion maps are obtained through spectral fitting using the Galaxy Integral field unit Spectroscopy Tool\footnote{\url{https://pypi.org/project/gistPipeline/}} \citep[{\sc gist},][]{Bittner2019}, which is a wrapper for the penalised pixel cross-correlation fitting {\sc python} package ({\sc pPXF}; \citealt{Cappellari2004,Cappellari2017}). The fitting is restricted to the first two velocity moments.

A reduced version of IndoUS is used for the stellar template library \citep{Valdes2004}. Spaxels within the half-light radius (derived from  {\sc ProFound}) are added to construct an aperture spectrum, whose fit provides the initial redshift estimate of the object. A subset of stellar templates is first selected for fitting individual spaxels, following the method of \citet{vandeSande2017}. For each galaxy, a series of concentric elliptical annuli is constructed, considering only spaxels with a signal-to-noise ratio S/N $>3$. The annuli are centred on the galaxy, with ellipticities defined using {\sc ProFound} axis ratios, and are iteratively expanded in steps of half a pixel until a target S/N of 25 is reached or no additional valid spaxels remain. The integrated spectrum within each annulus is then fitted, and the templates with non-zero weights are retained. Individual spaxels are subsequently fitted using a reduced set of stellar templates, consisting of those selected for the corresponding annulus and its adjacent annuli. Uncertainties on the spaxel measurements are estimated via Monte Carlo simulations, in which multiple realisations of the input spectra are generated by adding random noise based on the residuals of the fit.

\subsection{Ionised gas kinematics}
\label{Ionised gas kinematics}

To identify the ionised gas spin axes, we exploit the MAGPI emission line data product release presented by \citet{Battisti2026}. Briefly, {\sc gist}, which incorporates the "Gas and Absorption Line Fitting" package \citep[\textsc{GandALF};][]{Sarzi2006,Falcon2006}, is used to fit the observed (seeing-convolved) spectra of MAGPI galaxies. The stellar continuum is 
modelled on a spaxel-by-spaxel basis using the C3K spectral library \citep{Conroy2009}, based on the {\sc mist} isochrones \citep{Choi2016}, with emission lines masked during the fitting. This step enables an accurate subtraction of the stellar continuum prior to emission-line modelling. 

Ionised gas mean velocity and velocity dispersion maps are then derived using {\sc gist}, where emission lines are fit with single Gaussian components assuming a common kinematic solution across all lines, with uncertainties estimated via Monte Carlo realisations. Due to the MAGPI redshift range ($0.26\leq z \leq 0.42$) and the MUSE wavelength coverage (470--935 nm) only a subset of emission lines is detected with sufficient S/N $>3$ (see table~1 of \citealp{Battisti2026}). Thus, the resolved ionised gas kinematics is effectively constrained by the strongest available emission line with the highest S/N which is tracked via the \texttt{VTIELINE} header keyword. Table~\ref{EmissionLines} summarises the reference emission lines and approximate redshift windows up to $z$\,$\sim$\,1.5. Although the main aim of the MAGPI Survey is to study physical drivers of galaxy evolution at lookback times of 3--4 Gyr ($0.26\leq z\leq 0.42$), additional galaxies up to $z$\,$\sim$\,1.5 are serendipitously detected in the MUSE field-of-view and included in the emission line data product release. The exact redshift window(s) for each line depend on the galaxy kinematics and are determined by combining the redshift, line velocity, and sigma maps with the wavelength mask used in \textsc{GandALF} (see table~2 of \citealp{Battisti2026}).

\begin{table}
	\centering
	\caption{Reference emission lines for MAGPI ionised gas kinematics up to $z\sim1.5$. Wavelengths are in vacuum. The approximate redshift window(s) for each line are listed, where some have breaks due to the notch filter to block the GLAO sodium laser at 580--597 nm. The reference line used in this work.}
	\label{EmissionLines}
	\begin{tabular}{lcc}
		\toprule
		Line & $\lambda_\mathrm{rest}$ (\AA)  & Redshift window(s)\\
		\midrule
        $\mathrm{[O\,II]}$   & 3727.09  & $0.262-0.556;\,0.602-1.508$\\
        
        $\mathrm{[O\,II]}$   & 3729.79  & $0.261-0.555$; $0.601-1.507$\\ 
        
      %  %[O\,III]  & 4364.38  & $0.077-0.328$; $0.368-1.142$\\
        
      %  %[O\,III]  & 4960.21  & $0.000-0.169$; $0.204-0.885$\\
        
      $\mathrm{[O\,III]}$ & 5008.17  & $0.000-0.158$; $0.193-0.867$\\
       H$\alpha$ & 6564.61 & $0.000-0.424$\\
		\bottomrule
	\end{tabular}
\newline  
\end{table}

\subsection{Kinematic position angles}
\label{Kinematic position angles}

The stellar kinematic position angles (PA$_{\rm stars}$) are measured from the spatially-resolved stellar mean velocity maps using the {\sc fit\_kinematic\_pa} routine (see appendix~C of \citealp{Krajnovic2006}). Only spaxels with stellar continuum S/N$>$3 and velocity uncertainty $<$30\,km\,s$^{-1}$ are used in the fitting, following \citet{Bryant2019}. 

The ionised gas kinematic position angles (PA$_{\rm gas}$) are estimated using the spatially-resolved ionised gas mean velocity maps, which are tied to the reference emission line (H$\alpha$, [O\,III], or [O\,II] depending on the redshift windows (see Table~\ref{EmissionLines}). This reflects the varying wavelength coverage of MUSE across the MAGPI redshift range up to $z\sim 1.5$. We use the {\sc fit\_kinematic\_pa} routine accounting for only spaxels with S/N $>5$ \citep{Bryant2019} in the reference emission line. Since we focus on the redshift range $0.20<z<0.42$, for this work we select only PA$_{\rm gas}$ measurement from the reference emission line H$\alpha$ for tracing consistent ionised gas (see Table~\ref{EmissionLines}). 

\begin{figure}
\includegraphics[width=\columnwidth]{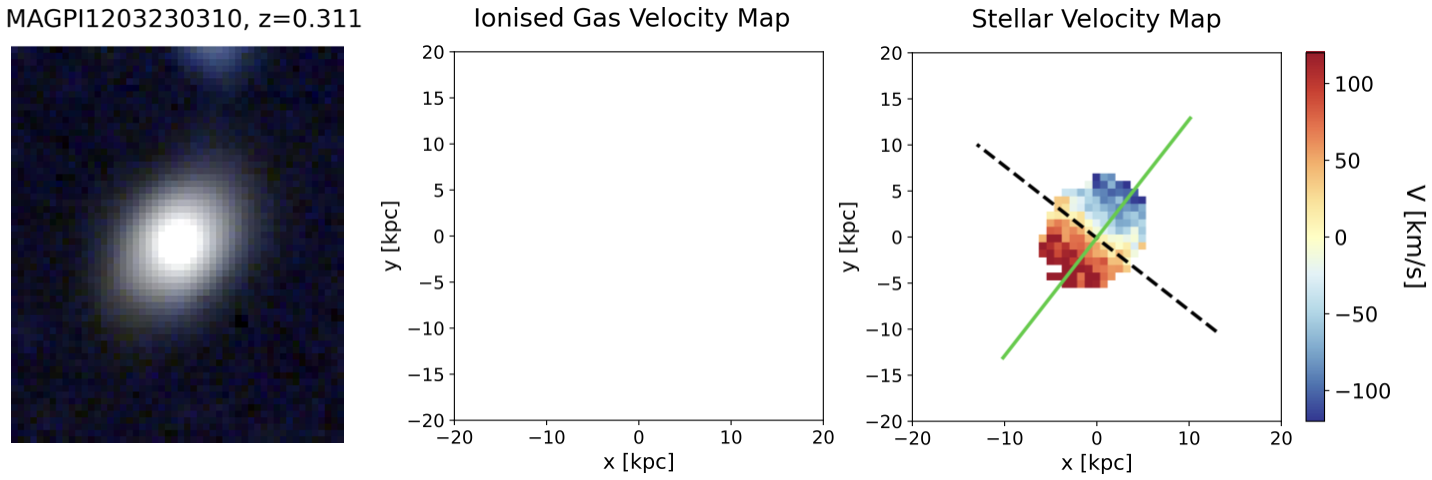} 
\includegraphics[width=\columnwidth]{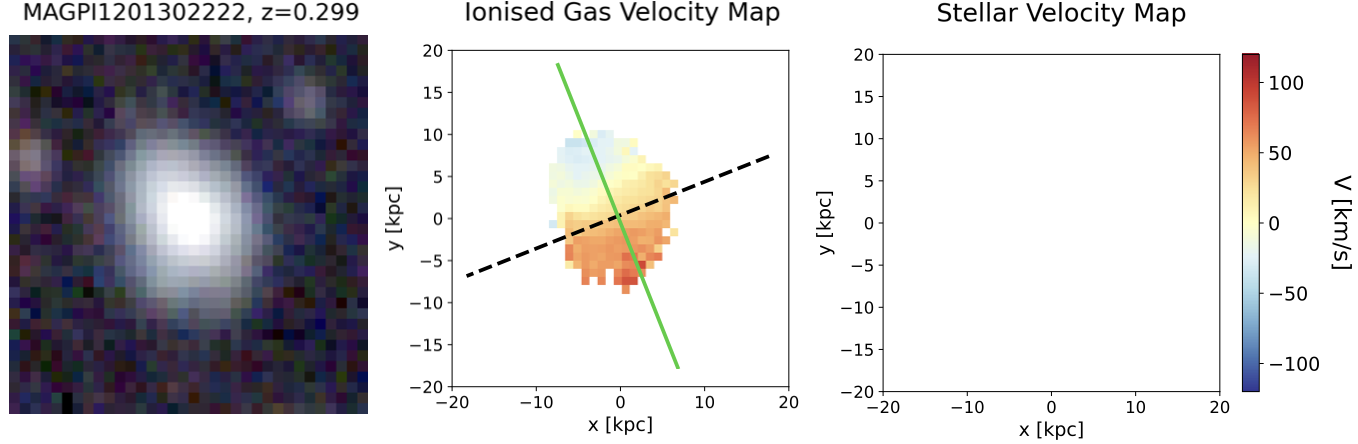} 
\includegraphics[width=\columnwidth]{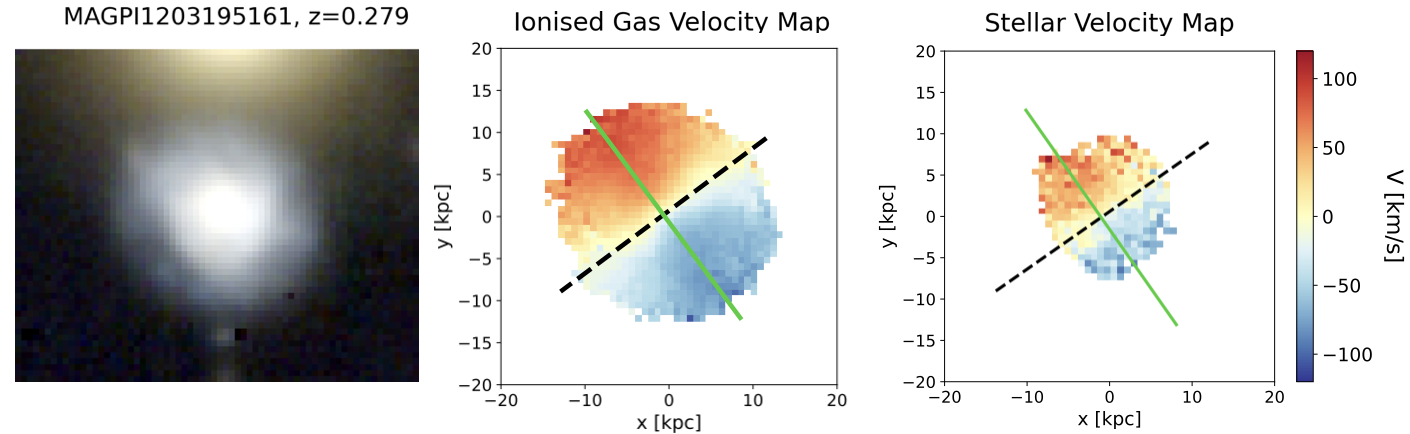}
\vspace*{-6pt}
\caption{Examples of kinematic PAs from stellar and ionised gas (H$\alpha$) mean velocity maps for MAGPI galaxies at $z\sim0.3$. Only spaxels meeting the quality criteria are included, resulting in galaxies with measurements of $\mathrm{PA}_{\rm stars}$ only (top panel), $\mathrm{PA}_{\rm gas}$ only (middle panel), or both (bottom panel). All are included in this study. The green line shows the kinematic PA and the black dashed line marks the projected axis of rotation. The left panels shows an RGB image of the MAGPI galaxies created from $g$ (blue), $r$ (green), and $i$ (red) from the MUSE datacube.}
\label{KinPAs_examples}
\end{figure}

Uncertainties on the kinematic position angles are estimated following \citet{Krajnovic2006}, by exploring the variation of the goodness-of-fit as a function of position angle and identifying the range corresponding to a $\Delta\chi^2$ confidence interval around the best-fitting solution. Figure~\ref{KinPAs_examples} shows examples of stellar and ionised gas mean velocity maps for MAGPI galaxies at $z\sim0.3$, where the green lines indicate the kinematic position angles.

\subsection{Stellar mass}
\label{Stellar mass}
Stellar masses for MAGPI galaxies are derived following the method developed for the Galaxy And Mass Assembly (GAMA; \citealp{Driver2011,Driver2022}) survey from \citet{Bellstedt2020} and based on the {\sc ProSpect} spectral energy distribution fitting code \citep{Robotham2020}. {\sc ProSpect} is used to derive stellar masses based on broad-band photometry in 9 bands (ugriZYJHKs) with a \citet{Chabrier2003} initial mass function. Forced photometry based on the segmentation maps is derived using images from the Kilo-Degree Survey (KiDS; \citealp{deJong2017}) and VISTA Kilo-degree Infrared Galaxy (VIKING; \citealp{Edge2013}) that are
pixel-matched to the MAGPI minicubes. We adopt the stellar mass defined as the median of the posterior distribution, with uncertainties given by half the 16--84 percentile range.

\subsection{Selection criteria}
\label{Selection criteria}

We focus on the redshift range $0.20<z<0.42$. Selecting galaxies with $z>0.2$ ensures a clear separation from local IFU surveys such as the SAMI Galaxy Survey (at $z<0.1$), enabling a cleaner comparison between low- and intermediate-redshift galaxy populations. The upper redshift limit is set by the visibility of H$\alpha$ within the wavelength coverage of MUSE, ensuring a consistent tracer of ionised gas across the sample. Moreover, beyond this redshift the reliability of the reconstructed cosmic web decreases, as the number density of available galaxies used in its reconstruction drops significantly. At $0.20<z<0.42$, there are 1443 galaxies with available minicubes and 722/1443 with measured stellar mass $7<\log(M_\star/M_{\odot})<12$. Beyond $z\sim0.4$ MAGPI galaxies are mostly unresolved.

\begin{figure*}
\centering
\includegraphics[width=\textwidth]{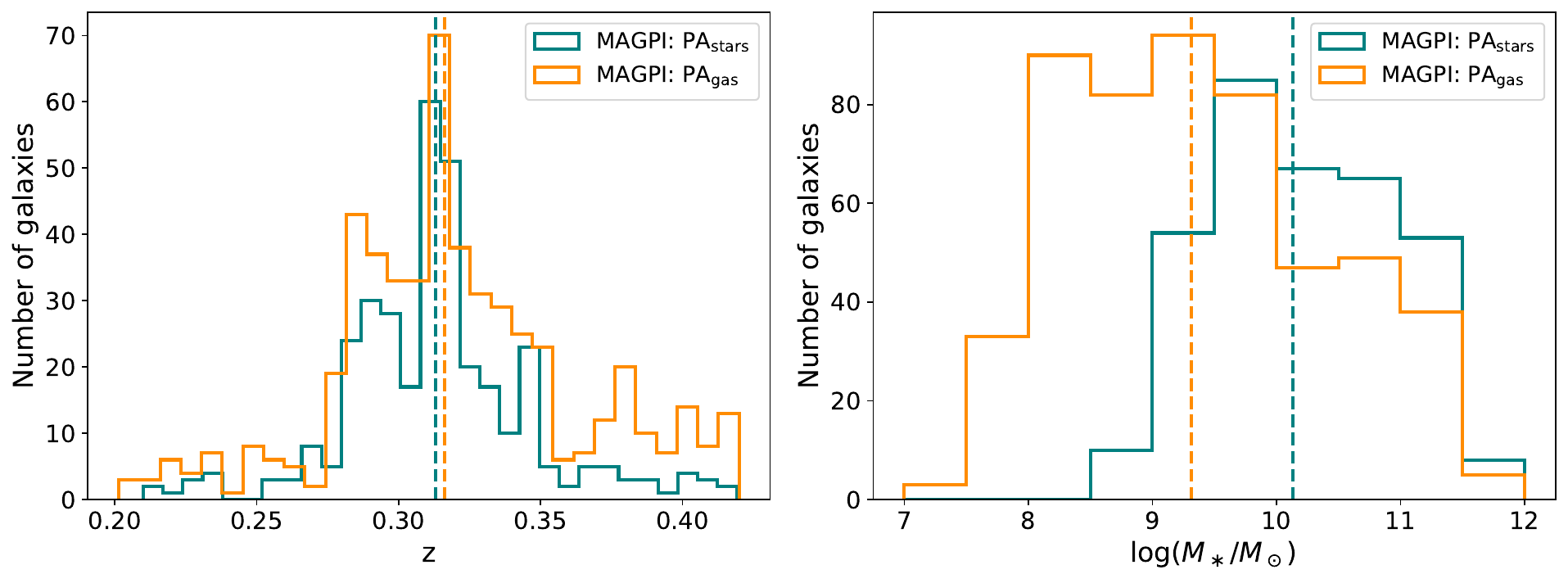}
\includegraphics[width=\textwidth]{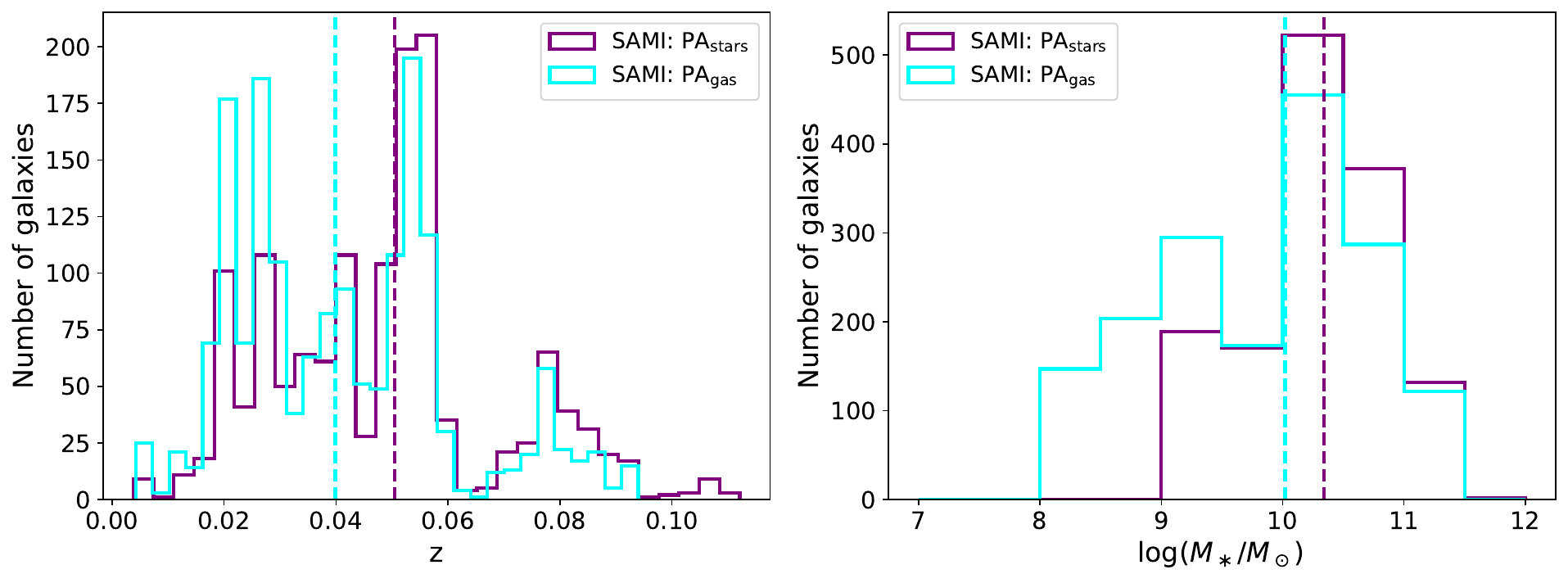}
\vspace*{-12pt}
\caption{Top panel: distributions of spectroscopic redshifts and stellar masses at $0.20<z<0.42$ for 342 MAGPI galaxies with measured $\rm PA_{stars}$ (teal) and 523 MAGPI galaxies with measured $\rm PA_{gas}$ (orange); dashed lines are medians. Bottom panel: distributions of spectroscopic redshifts and stellar masses at $0<z<0.12$ for 1388 SAMI galaxies with measured $\rm PA_{stars}$ (purple) and 1683 SAMI galaxies with measured $\rm PA_{gas}$ (orange)}
\label{MAGPIproperties}
\end{figure*}

Stellar kinematics are available for 787 MAGPI galaxies up to $z\sim1.5$, with 644 having $\rm PA_{stars}$ measurements with 1$\sigma$ uncertainties $\delta{\rm PA}<25$\degree. This cut allows us to select converged fits. Out of these, 342 galaxies are within $0.20<z<0.42$ and have a measured stellar mass $8.5<\log(M_\star/M_{\odot})<12$. The median values are $z=0.313$ and $\log(M_\star/M_{\odot}) = 10.13 ^{+0.10}_{-0.11}$. Ionised gas kinematics are available for 2462 galaxies up to $z\sim1.5$, with 1735 having reliable $\rm PA_{gas}$ values. Out of these, 523 galaxies are at $0.20<z<0.42$ and have measured stellar mass $7<\log(M_\star/M_{\odot})<12$. The median values are $z=0.316$ and $\log(M_\star/M_{\odot}) = 9.31 ^{+0.21}_{-0.18}$. Top panel of Figure~\ref{MAGPIproperties} shows the MAGPI redshift and stellar mass distributions.

There are 242 MAGPI galaxies with measured both $\rm PA_{stars}$ and $\rm PA_{gas}$ at $0.20<z<0.42$ and with stellar mass $8.5<\log(M_\star/M_{\odot})<12$. Most of these galaxies (60\%) have the stellar and gaseous components kinematically aligned with $\Delta {\rm PA} = |{\rm PA}_{\rm stars}-{\rm PA}_{\rm gas}| < 15^\circ$.

\subsection{SAMI galaxy spin--filament alignments at $z$\,$\sim$\,0}
\label{SAMI galaxy survey}

As a $z$\,$\sim$\,0 reference for galaxy spin--filament alignments based on spatially-resolved kinematics, we use the Sydney--AAO Multi-object Integral-field (SAMI) Galaxy Survey. The SAMI survey is composed of more than 3000 galaxies with stellar mass range $\log(M_{\star}/M_{\odot})=8$--12 and redshift range $0 <z\leq 0.12$ \citep{Bryant2015,Croom2021}. Most of the SAMI targets belong to the three equatorial fields GAMA G09, G12 and G15 of the Galaxy And Mass Assembly survey \citep{Driver2011}. Eight massive clusters were also observed to explore high galaxy density environments \citep{Owers2017}, however they are excluded from SAMI galaxy spin--filament alignment studies \citep{Welker2020,Barsanti2022,Barsanti2023}.

The SAMI spectrograph was mounted on the 3.9\,m Anglo-Australian Telescope \citep{Croom2012}. The instrument has 13 fused optical fibre bundles (hexabundles), each containing 61 fibres of 1.6$^{\prime\prime}$ diameter so that each integral field unit (IFU) has a 15$^{\prime\prime}$ diameter \citep{Bland2011,Bryant2014}. The SAMI fibres feed the two arms of the AAOmega spectrograph \citep{Sharp2006}. The SAMI Galaxy Survey uses the 580V grating in the blue arm, giving a resolving power of $R=1812$ and wavelength coverage of 3700--5700\,\AA, and the 1000R grating in the red arm, giving a resolving power of $R=4263$ over the range 6300--7400\,\AA. The median full-width-at-half-maximum values for each arm are FWHM$_{\rm blue}$=2.65\,\AA\ and FWHM$_{\rm red}$=1.61\,\AA\  \citep{vandeSande2017}. The data are reduced using the SAMI {\sc python} package \citep{Allen2014}, which uses the {\tt 2dfdr} package \citep{2015ascl.soft05015A}. A complete description of the data reduction from raw frames to datacubes can be found in  \citet{Sharp2015}, \citet{Allen2015}, \citet{Green2018} and \citet{Scott2018}. The final datacubes are characterised by a grid of 0.5$^{\prime\prime}\times0.5^{\prime\prime}$ spaxels, where the blue and red spectra have pixel scales of 1.05\,\AA\ and 0.60\,\AA\ respectively. 

We use the SAMI galaxy sample built by \citet{Barsanti2022} as $z$\,$\sim$\,0 reference. This sample is characterised by 1388 galaxies with measured ${\rm PA}_{\rm stars}$ with mass $9<\log(M_\star/M_{\odot})<12$ and 1683 galaxies with measured ${\rm PA}_{\rm gas}$ with mass $8<\log(M_\star/M_{\odot})<12$. The redshift range is $0<z<0.12$ with median $z=0.050$. Thus, the median redshifts of the MAGPI ($z\sim0.3$) and SAMI ($z\sim0.05$) surveys are separated by  $\sim3$\,Gyr of cosmic evolution. Bottom panel of Figure~\ref{MAGPIproperties} shows the SAMI redshift and stellar mass distributions.

The cosmic filaments are reconstructed from the Galaxy And Mass Assembly (GAMA; \citealp{Driver2011}) spectroscopic redshift survey for the G09, G12 and G15 regions. To ensure consistency with the MAGPI galaxy analysis since spin--filament alignments can depend on the characterisation of filaments \citep{GaneshaiahVeena2021}, we built the cosmic web and the measurement of spin--filament alignments using the same methods as described below.

\section{Reconstruction of the cosmic web}
\label{Reconstruction of the cosmic web}

 The cosmic web reconstruction is performed combining wide-area spectroscopic surveys at $z\sim0.3$: the GAMA survey, the MAGPI Redshift Survey and the DESI survey. We use redshifts in the Cosmic Microwave Background (CMB) frame to ensure a consistent cosmological reference frame across the combined samples. This correction removes the contribution of the observer’s peculiar motion with respect to the CMB and provides accurate distances for the reconstruction of the density field. The MAGPI galaxies are overlaid on the reconstructed cosmic web and assigned to their nearest cosmic filament for the spin--filament alignment analysis.

\subsection{The GAMA survey}
\label{GAMA galaxy survey}

In order to reconstruct the cosmic web, we take advantage of the Galaxy And Mass Assembly (GAMA; \citealp{Driver2011}) survey underlying the MAGPI Survey. GAMA is a spectroscopic and photometric survey of $\sim$300,000 galaxies down to $r < 19.8$\,mag that covers $\sim$286\,deg$^{2}$ in 5 regions called G02, G09, G12, G15 and G23. The redshift range of the GAMA sample is $0<z<0.5$, with a median value of $z\sim0.25$. Most of the spectroscopic data were obtained using the AAOmega multi-object spectrograph at the Anglo-Australian Telescope, although GAMA also incorporates previous spectroscopic surveys such as SDSS \citep{York2000}, 2dFGRS \citep{Colless2001,Colless2003}, WiggleZ \citep{Drinkwater2010} and the Millennium Galaxy Catalogue \citep{Driver2005}. The deep and highly complete spectroscopic redshift data (98.5\% in the equatorial regions; \citealp{Liske2015}), combined with the wide area, make the GAMA survey an ideal galaxy redshift sample for extracting the filaments of the cosmic web.

We make use of the DR4 data release of the GAMA survey \citep{Driver2022}. We select 88,481 galaxies with secure redshifts and stellar masses in the MAGPI redshift range ($0.20<z<0.42$) lying within the G12, G15 and G23 regions (where MAGPI galaxies have been observed). Galaxies have stellar masses $7<\log(M_{\star}/M_{\odot})<12$; the median stellar mass and redshift are $M_\star \sim 10^{10.73}\,M_{\odot}$ and $z=0.280$.

The GAMA magnitude limits ($r<19.8$ for G12/G15; $i<19.2$ for G23) maximise galaxy sampling at $z<0.2$. At higher redshift, the average galaxy number density drops from 8$\times10^{-3}$ Mpc$^{-3}$ to 2$\times10^{-3}$ Mpc$^{-3}$. Moreover, the average stellar mass increases from $\log(M_{\star}/M_{\odot})=9.5$ at $z<0.1$, to $\log(M_{\star}/M_{\odot})=10.3$ at $0.1<z<0.2$, and to $\log(M_{\star}/M_{\odot})=10.8$ at $0.2<z<0.42$, reflecting the progressive loss of low-mass galaxies in this flux-limited sample. The combination of reduced number density and mass-dependent incompleteness significantly impacts the reliability of cosmic web reconstruction from the GAMA survey alone at $z\sim0.3$: the sparse sampling increases shot noise in the inferred density field, while the preferential selection of massive galaxies leads to an incomplete tracing of low-density environments. As a result, filamentary structures may be artificially fragmented or missed altogether in the regions where MAGPI galaxies are embedded. As a consequence, we complement the GAMA sample with spectroscopic redshifts from the MAGPI Redshift Survey (Section~\ref{The MAGPI Redshift Survey}) and the DESI survey (Section~\ref{DESI DR 1 survey}) in order to increase the number density of tracers and improve the fidelity of the reconstructed large-scale structure.

\subsection{The MAGPI Redshift Survey}
\label{The MAGPI Redshift Survey}

\begin{figure*}
\centering
\includegraphics[width=\textwidth]{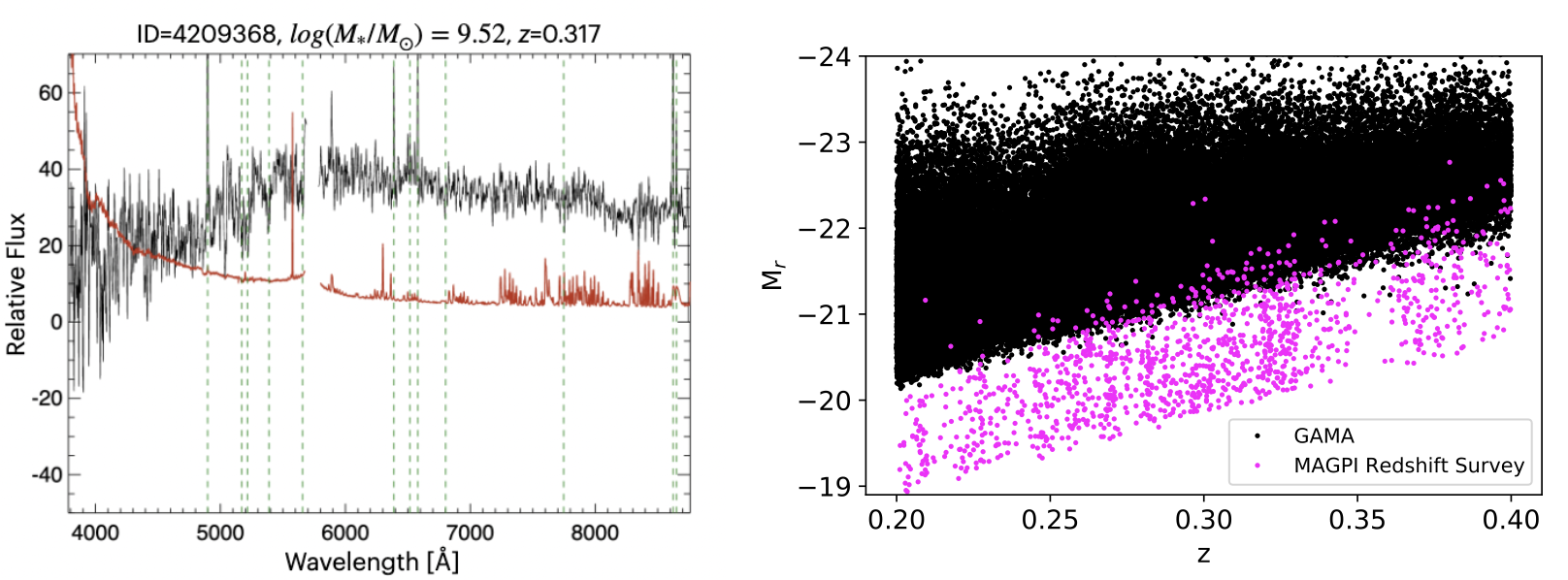}
\vspace*{-12pt}
\caption{Left panel: an example spectrum for a galaxy of the MAGPI Redshift Survey with the red line showing the variance. Right panel: $r$-band absolute magnitude as a function of redshift for the MAGPI Redshift Survey (magenta) and the GAMA survey (black).}
\label{MRSplots}
\end{figure*}

To enhance the reconstruction of the cosmic web at $z\sim0.3$, we designed the MAGPI Redshift Survey (P.I. Barsanti) carried out using the 2dF multi-object spectrograph mounted on the Anglo-Australian Telescope. The MAGPI Redshift Survey aims to obtain spectroscopic redshifts for galaxies down to $r<21$ in G12 and G15, and $i<20.5$ in G23, within a projected radius of 5\,Mpc around each MAGPI central galaxy. This strategy enables us to probe lower mass galaxies and trace the large-scale environment on Mpc scales at look-back times of 3--4\,Gyrs. The survey targets objects $\sim$1.2 magnitudes deeper than GAMA, corresponding to a factor of $\sim$3 increase in sampling depth, and aims to reach a spectroscopic completeness $>90\%$ for galaxies with $\log(M_{\star}/M_{\odot})>9.5$ at $z\sim0.3$. The 5\,Mpc radius allows us to trace the full extent of haloes and minimise boundary effects. It also provides sufficient volume to robustly identify filaments, whose typical widths are of order $\sim$1\,Mpc, around the MAGPI galaxies.

During 2023 this project was allocated 6.5 nights. We used the lowest resolution setup of AAOmega, the 580V and 385R gratings, which have a resolution of R$\sim$1000 (blue) to R$\sim$1600 (red) and a wavelength coverage of 370–880 nm. We derive an exposure time of 1.2 hours at medium seeing 1.5$\arcsec$ to reach S/N=3\,\AA$^{-1}$ and S/N=4.2\,\AA$^{-1}$ for the blue and red arms, respectively.

Data were successfully observed, reduced \citep{Owers2017} and analysed for 6200 galaxies around 13 out of 56 MAGPI central galaxies in G12, G15 and G23. Out of 6200 galaxies we measure reliable spectroscopic redshifts for 3720 galaxies (60\%) with a probability of being correct $>$\,0.9 using {\sc autoz}, the automated software tool developed for the GAMA survey to determine galaxy redshifts from spectroscopic data \citep{Baldry2014,Owers2017}. The left panel of Figure~\ref{MRSplots} shows an example spectrum. We find that 1317 galaxies have $0.2<z<0.42$ and $8.9<\log(M_{\star}/M_{\odot})<11.3$. The median values are $z\sim0.300$ and $M_\star \sim 10^{10.14}\,M_{\odot}$. 

{\sc ProSpect} is used to measure stellar mass with a procedure similar to that used for MAGPI galaxies (Section~\ref{Stellar mass}). The right panel of Figure~\ref{MRSplots} shows the $r$-band absolute magnitude ($M_r$) as a function of redshift for a pseudo-volume-limited population of galaxies combining the GAMA survey and the MAGPI Reshift Survey. This population is defined as all galaxies with $M_r(z_{\rm ref}$=0, $Q$=0.78) $<-21.2$\,mag, where $Q$ accounts for the expected evolution of $M_r$ with redshift \citep{Loveday2015}. The inclusion of the MAGPI Redshift Survey extends the galaxy population to significantly fainter magnitudes at fixed redshift. By construction, there are no duplicate galaxies between GAMA and the MAGPI Redshift Survey. We explain the lack of galaxies at $z\sim0.35$ in the next Section.

%Since 2024, the 2dF multi-object spectrograph has been offline for upgrades and refurbishment. We plan to resume the MAGPI Redshift Survey in 2027, targeting the environments of the remaining 43 MAGPI central galaxies. This will complete the MAGPI Redshift Survey to its full design, reaching a total sample of $\sim$34,825 galaxies (excluding duplicates from public spectroscopic redshift surveys) across all 56 MAGPI centrals. The full survey requires approximately 307 hours of observing time over 37 nights, including all overheads.

\subsection{The DESI survey}
\label{DESI DR 1 survey}

To improve the sampling of the large-scale structure, we complement the GAMA galaxy sample and the MAGPI Redshift Survey with spectroscopic data from the Dark Energy Spectroscopic Instrument Data Release 1 (DESI DR1; \citealp{Karim2026}). DESI DR1 provides spectroscopic redshifts for 13.1 million galaxies at $0<z<4$. Redshifts in DESI are derived from high-throughput optical spectroscopy obtained with a multi-object fibre spectrograph mounted on the 4m Mayall Telescope in Arizona, United States. DR1 consists of all data acquired during the first 13 months of the DESI main survey, from May 2021.

We construct the spectroscopic galaxy sample from DESI DR1 using the {\sc zpix} redshift catalogue. We select objects classified as reliable galaxy redshifts by requiring spectype = "GALAXY", zwarn = 0, and zcat\_primary = "T", ensuring a clean primary redshift solution with high confidence. The sample is further restricted to the GAMA regions G12, G15 and G23, where MAGPI galaxies have been observed, and to $0.20<z<0.42$. We find 44,198 galaxies in G12, 39,409 in G15 and 0 in G23 (this last region is not covered by DESI). We cross-match the combined sample of 83,607 galaxies with the value-added catalogue {\sc Mass EMLines} for extragalactic science that provides stellar mass estimates from CIGALE \citep{Zou2024}. These estimates are based on the broad-band g, r, z, W1, and W2 photometry from the DESI Legacy Imaging Surveys, and spectrophotometry of 10 artificial bands generated through convolution with DESI spectra. The resulting sample has 83,589 galaxies with $7<\log(M_\star/M_{\odot})<12$.

We search for duplicate objects in the DESI catalogue by cross-matching with GAMA and the MAGPI Redshift Survey within a 1$\arcsec$ radius, identifying 47,108 matches between GAMA and DESI and 256 matches between DESI and the MAGPI Redshift Survey. The duplicate galaxies are used to assess the consistency between the CIGALE stellar mass estimates for DESI and the ProSpect estimates for GAMA and the MAGPI Redshift Survey. For the 47,108 galaxies common to DESI and GAMA, the DESI masses tend to be higher by a median of $0.118$ dex, with a scatter of $0.165$ dex. The central $68\%$ of the mass differences spans approximately $-0.003$ to $0.264$ dex, indicating that the two estimates are consistent within a few tenths of a dex. For the 256 galaxies common to DESI and the MAGPI Redshift Survey, the agreement is even closer, with a median offset of $-0.036$ dex and a scatter of $0.122$ dex, both of which are comparable to the uncertainties on individual mass estimates. These comparisons indicate that the CIGALE and ProSpect stellar mass estimates are consistent for combining the three redshift samples.

\begin{figure}
\centering
\includegraphics[width=\columnwidth]{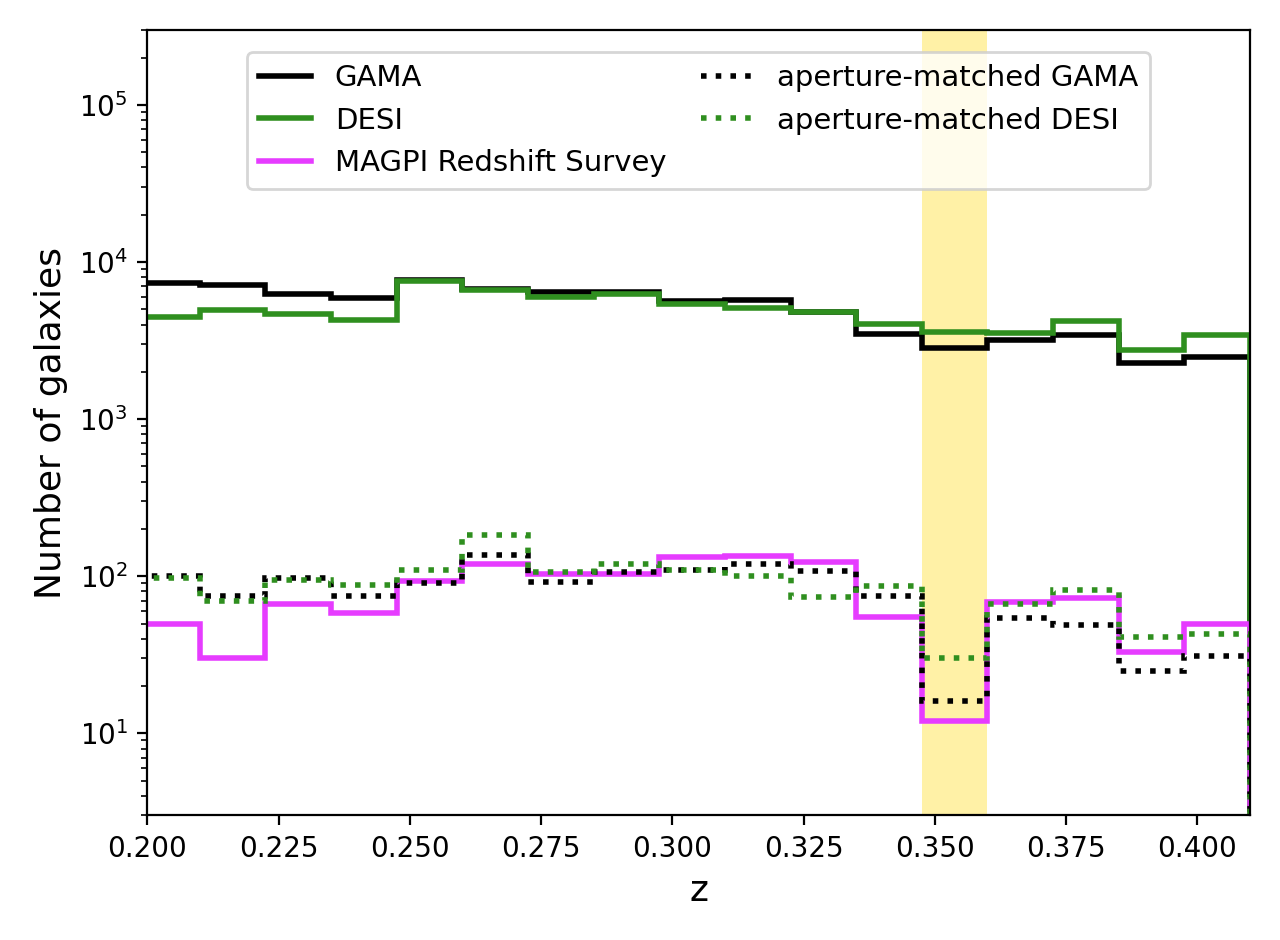}
\vspace*{-12pt}
\caption{Spectroscopic redshift distributions for the selected GAMA (black), MAGPI Redshift Survey (magenta) and DESI (green) galaxies. The dotted lines show GAMA and DESI galaxies selected within 5 Mpc of the 13 MAGPI central galaxies observed in the MAGPI Redshift Survey (aperture-matched). The yellow band marks the bin $0.3475 < z < 0.36$, where the MAGPI Redshift Survey shows a deficit of galaxies that is also seen in the aperture-matched GAMA and DESI samples.}
\label{z_distributions_GAMA_MRS_DESI}
\end{figure}

These duplicates galaxies are removed from DESI to ensure a unique galaxy sample. The resulting clean DESI sample contains 36,463 galaxies, with a median stellar mass of $\log(M_\star/M_\odot) \sim 10.69$ and median redshift $z \sim 0.295$. Figure~\ref{z_distributions_GAMA_MRS_DESI} shows the spectroscopic redshift distributions for the selected GAMA, MAGPI Redshift Survey and DESI galaxies, illustrating the $z$ ranges at which the DESI and MAGPI Redshift surveys are complementing GAMA. The lack of galaxies at $z\sim0.35$ in the MAGPI Redshift Survey is also present in the GAMA and DESI redshift distributions within the same 5 Mpc aperture for the 13 MAGPI centrals observed by the MAGPI Redshift Survey, with approximately four times fewer galaxies than the same surveys show at the same redshift outside the 5 Mpc apertures. This suggests that the deficit is unlikely to arise solely from incompleteness in the MAGPI Redshift Survey and may instead reflect the underlying large-scale structure, potentially including an underdense region along these lines-of-sight.

\subsection{Tracer populations: the Combined Redshift Sample}
\label{Tracer populations: the combined redshift sample}

We perform the cosmic web reconstruction using as galaxy tracers the Combined Redshift Sample (CRS), constructed from the GAMA survey, the MAGPI Redshift Survey, and the DESI survey, excluding duplicates (Section~\ref{DESI DR 1 survey}). The resulting sample contains 126,261 galaxies, with a median redshift of $z \sim 0.284$ and a median stellar mass of $\log(M_\star/M_\odot) \sim 10.71$. There are 47,089 galaxies in G12, 55,627 in G15 and 23,545 in G23.

\begin{figure}
\includegraphics[width=\columnwidth]{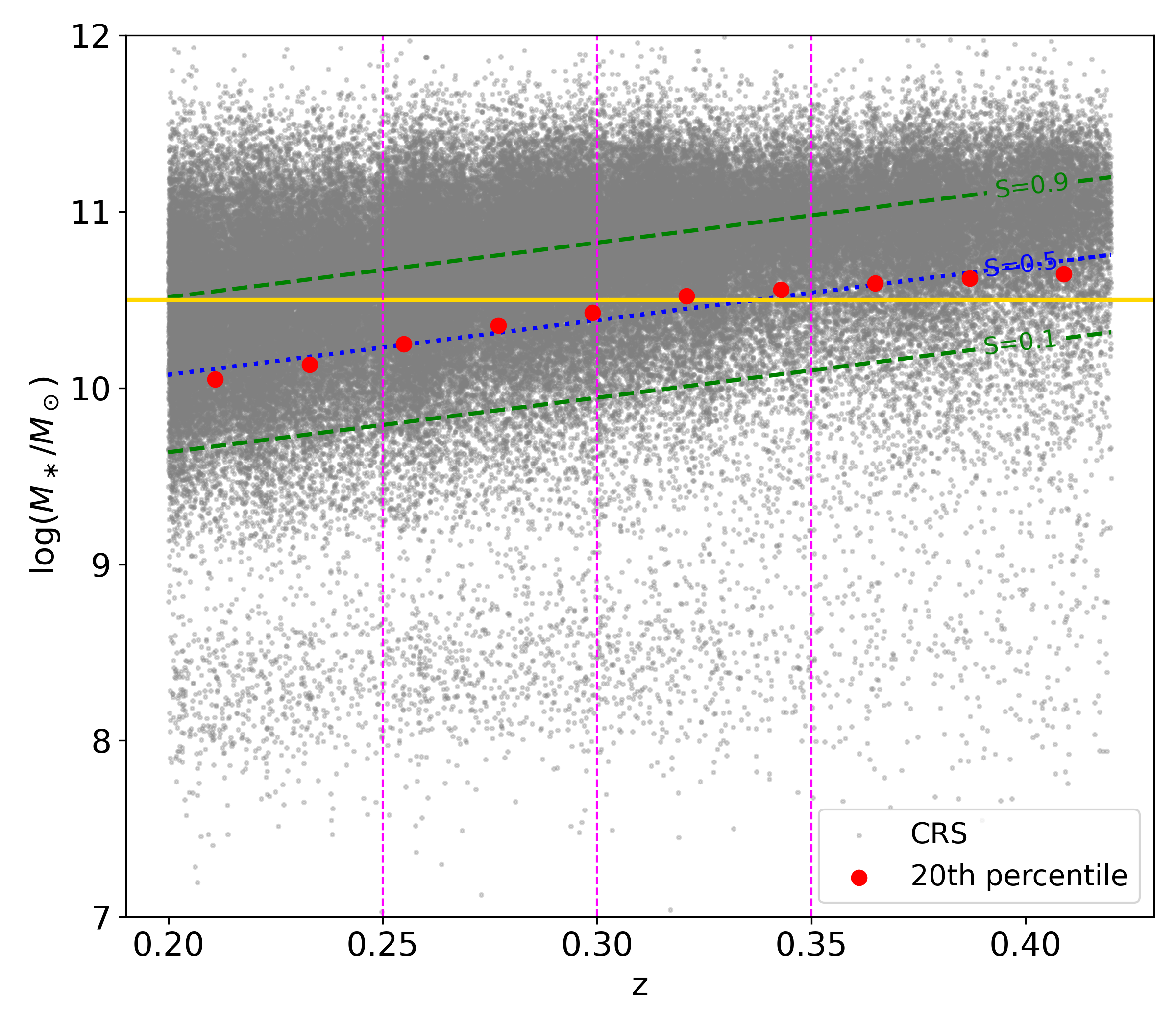}
\vspace*{-18pt}
\caption{Stellar mass versus spectroscopic redshift for the CRS (grey points). The yellow solid line marks the fixed mass cut $\log (M_{\star}/M_{\odot}) \geq 10.5$. The blue dotted line marks the evolving mass cut $\log M_{\star,lim}(z)=9.46+3.10z$ fitted to the 20th percentile masses (red points) and corresponding to $S=0.5$, the midpoint of the adopted probabilistic selection function $S(z,\log M_{\star})$. The green lines are $S=0.1$ and $S=0.9$, respectively, illustrating the width of the transition in the selection function. The magenta dashed lines mark the redshift bins for the mass-matched constant-number-density method.}
\label{logMstar_z}
\end{figure}

Figure~\ref{logMstar_z} shows stellar mass as a function of redshift for the CRS. Low-mass galaxies become progressively undersampled at higher redshift. This trend reflects the flux-limited nature of the contributing surveys, where only the most luminous (and therefore typically more massive) galaxies are observable at larger distances. As a result, the lower envelope of the stellar mass distribution increases with redshift, indicating a redshift-dependent mass completeness limit.

This incompleteness can have important implications for the reconstruction of the cosmic web. Massive galaxies are more strongly clustered and trace the underlying density field and filamentary structure more effectively than low-mass galaxies. Using simulations at $z$\,$\sim$\,0, \citet{Zakharova2023} showed that extending the sample to lower stellar masses does not significantly improve cosmic web reconstruction, as low-mass galaxies are more uniformly distributed across environments and therefore provide relatively weak constraints on the filamentary network. Nevertheless, the decreasing number density of low-mass galaxies at higher redshift can introduce artificial gradients in the tracer distribution, leading to a redshift-dependent bias in the inferred density field. Consequently, filamentary structures may appear less prominent or be under-detected at higher redshift, while being better resolved at lower redshift where the sampling is denser. 

\begin{table*}
\centering
\begin{tabular}{lcccccc}
\toprule
Method & $N$ & $f_{\rm cut}(\%)$ & $\sigma_{n}/\overline{n}$ & $d$\,(Mpc) & $n_{\rm eff}$ & JSD\\
\midrule
Full sample    & 126,261 &  0  & 0.625  & 8.130 & 126,261 & 0.053\\
Fixed $M_{\star}$ cut &   86,587   & 31 & 0.437 & 9.218 & 86,587 & 0.028\\
Redshift-dependent $M_{\star}$ cut  & 101,302  & 20 & 0.607 & 8.748 & 101,302 & 0.099 \\
Weights & 126,261  & 0 & 0.625 & 8.130 & 86,202 & 0.053 \\
Constant number density & 61,400  &  50 & 0.364 & 10.337 & 61,400 & 0.060\\
\textbf{Mass-matched constant-number-density} & 88,485  & 30 & 0.366 & 9.152 & 88,485 & 0.011\\
\bottomrule
\end{tabular}
\caption{Comparison of different selection strategies for defining the galaxy tracer population used for cosmic web reconstruction from the CRS. Column~1 lists the method, column~2 the number of galaxies, column~3 the percentage of removed galaxies, column~4 the variation in number density with redshift, column~5 the mean inter-galaxy separation, column~6 the effective number density and column~7 the Jensen–Shannon divergence. The bold text highlights the fiducial method used in this work.}
\label{tab:selection_methods}
\end{table*}

To mitigate these effects, it is necessary to account for the evolving mass completeness of the sample when defining the galaxy population used for cosmic web reconstruction. In this work, we therefore explore several selection strategies designed to achieve a more uniform sampling of the density field across the full redshift range, in order to minimise observational biases. We test six different approaches:

\begin{enumerate}
    \item  \textbf{Full sample.} This maximises the statistical power and preserves the highest possible tracer density for the reconstruction of the density field and filamentary structure. However, the sample is affected by redshift-dependent incompleteness, as low-mass galaxies are progressively missing at higher redshift. 
    
    \item \textbf{Fixed stellar mass cut.} This applies a constant stellar mass threshold $\log (M_{\star}/M_{\odot}) \geq 10.5$ (yellow line in Figure~\ref{logMstar_z}), reducing the sample to 86,587 galaxies (fraction of removed galaxies is $f_{\rm cut}\sim31$\%). This choice is motivated by the mass value above which the sample is largely complete across the full redshift range. The resulting sample provides a more uniform sampling of the density field and mitigates redshift-dependent selection effects. However, this approach removes a substantial fraction of low-redshift galaxies, thereby reducing the tracer density and degrading the spatial resolution of the reconstructed filamentary network.
    
    \item \textbf{Redshift-dependent stellar mass cut.} To better match the evolving completeness of the survey, we derive an empirical stellar mass limit as a function of redshift. Since the parent sample is effectively flux-limited, the minimum observable stellar mass increases with redshift. We therefore bin the CRS in redshift and compute the 20th percentile of the stellar mass distribution in each bin (red points in Figure~\ref{logMstar_z}), which traces the lower envelope of the observed population while remaining robust to outliers and low-number statistics. This is well-approximated by a linear relation, given by
    \begin{equation}
    \log M_{\star,{\rm lim}}(z)=a+bz 
    \end{equation}
    with $a=9.46$ and $b=3.10$ (blue line in Figure~\ref{logMstar_z}). Applying this redshift-dependent cut yields a sample of 101,302 galaxies ($f_{\rm cut}\sim20$\%), preserving more galaxies than the fixed cut while ensuring a more uniform level of completeness across redshift than the full sample.

    \item \textbf{Weights.} We retain the full sample and account for incompleteness by weighting galaxies in the density field reconstruction. This approach is conceptually similar to the method used by \citet{Kraljic2026} in reconstructing the cosmic web from only H$\alpha$ emitters detected by the Euclid satellite at $0.4<z<1.8$, where the biases are mitigated by applying stellar mass weighting. In our case, galaxies are assigned weights inversely proportional to their sampling probability, which depends on stellar mass and redshift, according to
    \begin{equation}
        w(z,\log M_{\star})=\frac{1}{S(z,\log M_{\star})}
    \end{equation}
    where $S(z,\log M_{\star})$ is the selection function (i.e.\ the probability that a galaxy of given $(z, \log M_{\star})$ is in the CRS). We model the selection function as a smooth function of stellar mass and redshift based on the empirically derived mass completeness limit. The latter is the redshift-dependent stellar mass cut obtained for method (iii), which provides an approximation to the survey completeness boundary. In this approach, galaxies in under-represented regions of the stellar mass-redshift plane are statistically up-weighted to compensate for the incompleteness of the CRS. A logistic form is adopted to ensure a smooth transition across the completeness boundary and to avoid introducing artificial features in the reconstructed density field, so that
    \begin{equation}
        S(z,\log M_{\star})=\frac{1}{1+exp[-\frac{\log M_{\star}-\log M_{\star,lim}(z)}{\Delta}]}
    \end{equation}
    with width parameter $\Delta=0.2$\,dex. The contours of this probabilistic selection function corresponding to $S=0.1$ (green), 0.5 (blue), 0.9 (green) are shown in Figure~\ref{logMstar_z}. Galaxies well above the completeness limit have selection probabilities approaching unity ($S\rightarrow1$) and therefore receive weights close to one, implying little or no incompleteness correction. Conversely, galaxies below the completeness boundary have progressively lower selection probabilities ($S\ll1$) and larger weights. At the completeness boundary itself, where $\log M_{\star}=\log M_{\star,\mathrm{lim}}(z)$, the selection function yields $S=0.5$, corresponding to a weight of $w=2$. To prevent highly incomplete regions of parameter space from dominating the density reconstruction and amplifying shot noise, the maximum allowed weights are capped at $w_{\rm max}=5$.
    
    \item \textbf{Constant number density.} As an alternative to stellar mass cuts or incompleteness weighting, we construct a sample with approximately constant comoving number density across redshift. In this approach, galaxies are ranked by stellar mass within individual redshift bins, and only the most massive galaxies are retained such that the tracer number density remains approximately uniform over the full redshift range. Unlike a fixed stellar mass cut, the method preserves a homogeneous sampling of the density field. Compared to the redshift-dependent stellar mass cut, the constant number density selection is less directly tied to the physical survey selection function. However, since this approach reduces the low-$z$ sample to match the sparse high-$z$ regime, the tracer population is substantially reduced to 61,400 ($f_{\rm cut}\sim50$\%), affecting the spatial resolution of the filamentary network.

    \item \textbf{Mass-matched constant-number-density.} We construct a redshift-matched tracer population to minimise redshift-dependent selection effects in the cosmic web reconstruction. The full sample is divided into four redshift intervals (magenta lines in Figure~\ref{logMstar_z}): $0.20 \leq z < 0.25$, $0.25 \leq z < 0.30$, $0.30 \leq z < 0.35$, and $0.35 \leq z \leq 0.42$. Stellar mass distributions are computed using bins of width 0.25 dex over the range $7.0 \leq \log(M_{\star}/M_{\odot}) < 12.5$. We adopt the highest-redshift interval, $0.35 \leq z \leq 0.42$, as the reference sample, since it represents the most restrictive selection and therefore defines the effective mass completeness limit of the CRS. For each lower-redshift bin, galaxies are randomly subsampled in stellar mass bins to reproduce the stellar mass distribution of the reference sample. This procedure yields a tracer population ($f_{\rm cut}\sim30\%$) with an approximately constant galaxy number density per unit comoving volume across redshift while preserving a consistent stellar mass distribution shape. The resulting matched sample reduces biases associated with redshift-dependent incompleteness and varying tracer populations.
\end{enumerate}
 
For all these strategies, we correct for the "Finger-of-God" effect (FoG; \citealp{Jackson1972}). This distortion effect is due to the random motions of galaxies within virialised large groups, and causes the elongation of halos in redshift space, possibly leading to the identification of spurious filaments. Following \citet{Kraljic2018}, we implement a compression correction by making galaxies isotropically distributed around their group centres. Groups are identified using the Friends-of-Friends code Nessie \citep{Lambert2026}, a RUST implementation of the original code used for GAMA group finding \citep{Robotham2011}. Nessie is run on the CRS and then links between MAGPI galaxies and CRS groups have been built using the Parliament code (Harborne et al., in prep.). There are 87 groups with at least 5 members in $0.2<z<0.42$ and there are 28,360 filaments for the mass-matched constant-number-density method.

Table~\ref{tab:selection_methods} shows a quantitative comparison of these selection strategies for the tracer populations of the cosmic web from the CRS, using the following metrics. For each approach, we quantify the completeness bias as the variation in number density with redshift, $\sigma_{n}/\overline{n}$, where $\overline{n}$ is the mean number density and $\sigma_{n}$ is the scatter across redshift bins. Small values of $\sigma_{n}/\overline{n}$ correspond to uniform completeness, while larger values indicate stronger selection bias. We quantify the impact on the density field sampling by computing the mean inter-galaxy separation, $d=n^{-3}$. Smaller $d$ values correspond to denser sampling and thus higher filament resolution. We use the effective number density, $n_{\rm eff}=(\Sigma w_i)^2/\Sigma w_i^2$, to quantify the effective number of statistically independent galaxies contributing to the density field reconstruction. For the full sample, the fixed $M_{\star}$ cut, the redshift-dependent $M_{\star}$ cut, the constant number density and the mass-matched constant-number-density samples, all galaxies are assigned uniform weights ($w_i=1$), so $n_{\rm eff}=N$, where $N$ is the total number of galaxies in the corresponding sample. In contrast, for the weighting-based approach, the actual weights are used in the computation of $n_{\rm eff}$. In this case, galaxies with larger weights contribute more strongly to the reconstructed density field but also increase the statistical variance, leading to an effective number density lower than the total number of galaxies. Finally, we quantify the stellar mass distribution stability across redshift using the Jensen–Shannon divergence (JSD, \citealp{Nielsen2020}) between redshift bins, testing whether the galaxy tracer population remains statistically comparable across redshift and is affected by evolutionary trends. JSD is a symmetric measure of the difference between probability distributions, where $\mathrm{JSD}=0$ corresponds to identical distributions (perfect redshift-independent tracer population) and larger values indicate increasing residual redshift-dependent population variation.

The mass-matched constant-number-density provides the best compromise between sample homogeneity and spatial sampling, reducing the variation in number density from $\sigma_{n}/\overline{n}=0.625$ for the full sample to 0.366 while retaining $\sim$70\% of the galaxies and maintaining a relatively small mean inter-galaxy separation ($d=9.15$\,Mpc). The fixed stellar mass cut achieves similar improvements, but it does so at the cost of a non-uniform stellar mass distribution across redshift (JSD$=0.028$) with respect to the mass-matched constant-number-density approach ($\mathrm{JSD}=0.011$),  corresponding to a $\sim60\%$ larger residual population variation. The redshift-dependent stellar mass cut preserves a larger fraction of galaxies but only minimally improves the uniformity of the tracer population ($\sigma_{n}/\overline{n}=0.607$) and is strongly affected by residual redshift-dependency ($\mathrm{JSD}=0.099$). The constant number density sample achieves the most homogeneous sampling ($\sigma_{n}/\overline{n}=0.364$) at the expense of a substantial reduction in galaxy number and increased tracer separation ($d=10.3$\,Mpc). The weighting-based approach yields an effective number density comparable to the mass-matched constant-number-density ($n_{\rm eff} \sim 86\times10^{4}$), indicating that the empirically derived selection function introduces only moderate corrections concentrated in sparsely sampled regions of the stellar mass--redshift plane. 

We therefore adopt the mass-matched constant-number-density as our fiducial tracer selection for this work, characterised by a highly redshift-invariant stellar mass distribution that minimises population-driven systematics in the spin–filament alignment analysis. Nevertheless, similar alignment trends are recovered using the fixed $M_{\star}$ cut, the full sample, and the weights approach. Appendix~\ref{Impact of cosmic web tracer selection on spin--filament alignments} shows that the presence of a transition stellar mass remains robust across these methods, with only modest insignificant shifts ($\sim$0.1–0.2 dex at $<1\sigma$ level) in the measured values depending on the adopted tracer selection strategy. On the other hand, the redshift-dependent $M_{\star}$ cut and the constant number density samples show larger residual variations in tracer population consistency (larger JSD values), poor density uniformity or larger separations; they perform poorly in reconstructing the cosmic web and are therefore excluded.

\subsection{Filamentary network}
\label{Filamentary network}

\begin{figure*}
\includegraphics[width=0.8\textwidth]{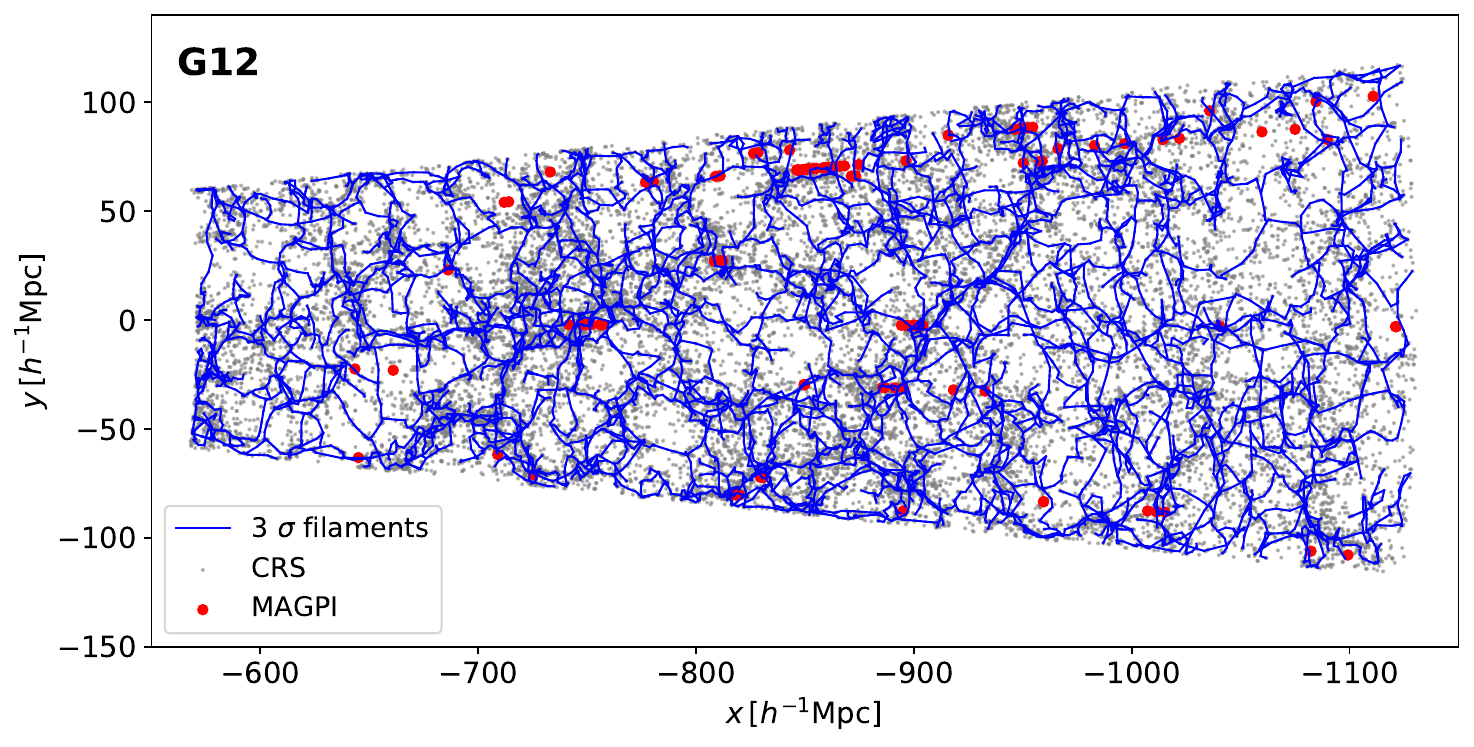}
\vspace{0.3cm} \\
\includegraphics[width=0.8\textwidth]{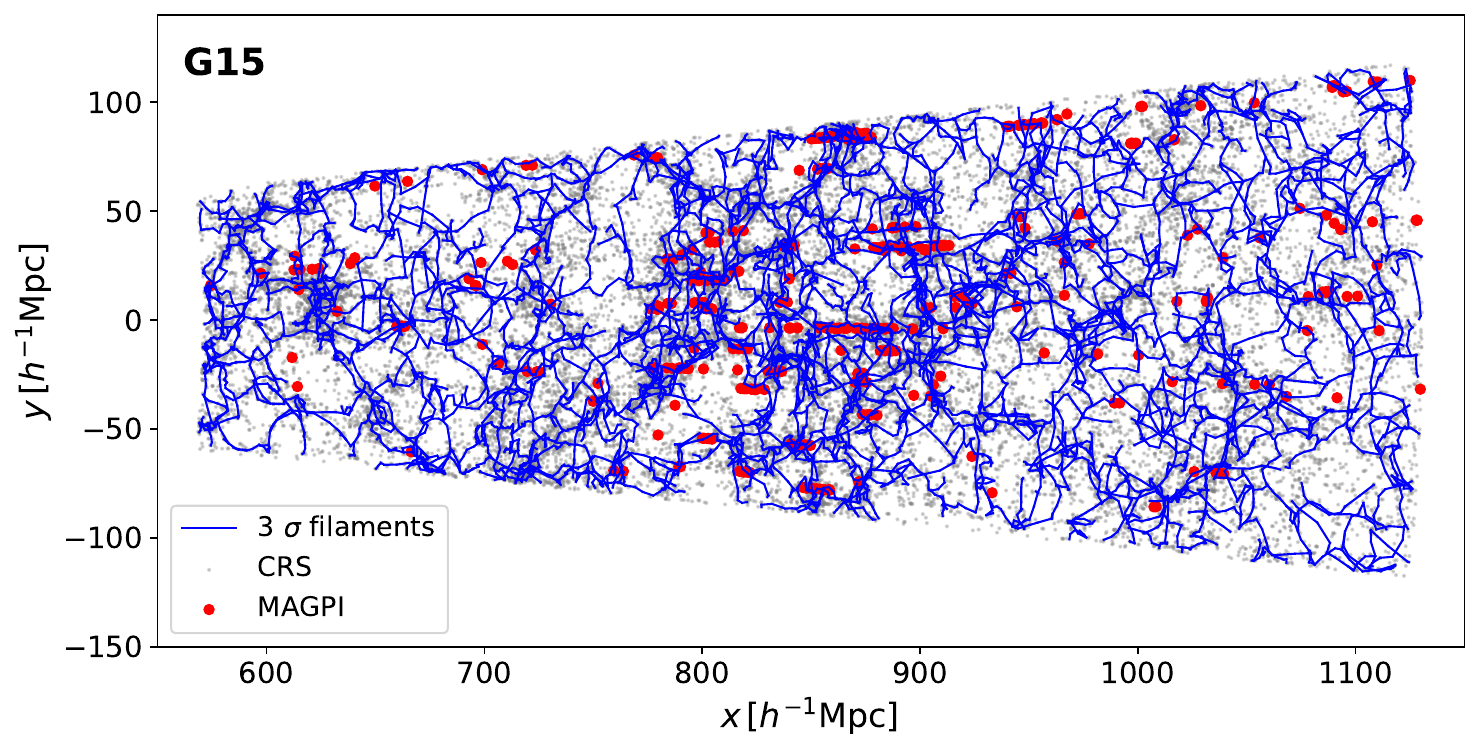}
\vspace{0.3cm} \\
\includegraphics[width=0.8\textwidth]{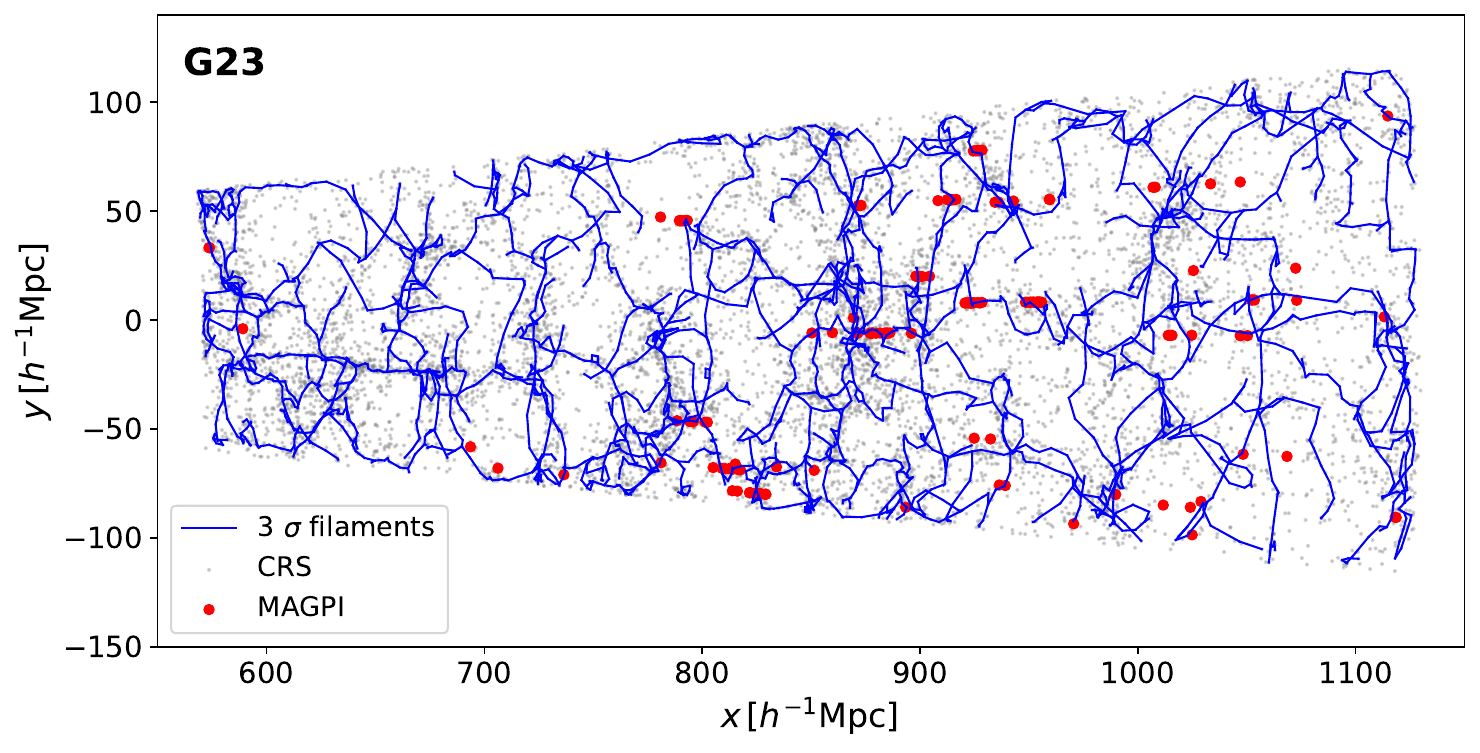}
\caption{Projected network of cosmic filaments (blue lines) for the G12, G15 and G23 regions at $0.20<z<0.42$. Grey points are 88,485 galaxies from the combined redshift sample used to reconstruct the spine of the cosmic web, applying the mass-matched constant-number-density selection for the tracer population chosen as the fiducial approach. Red points are the 722 MAGPI galaxies with available stellar mass measurements.}
\label{CosmicWeb}
\end{figure*}

The filaments are reconstructed using the Discrete Persistent Structure Extractor public code ({\sc DisPerSe}; \citealp{Sousbie2011a,Sousbie2011b}). {\sc DisPerSe} has been widely used in the literature to map the cosmic web, both using simulations (e.g., \citealp{Dubois2014,Welker2018,Codis2018}) and spectroscopic data, including the GAMA survey (e.g., \citealp{Barsanti2026,Barsanti2022,Kraljic2018, Duckworth2019, Welker2020}). Its strength for astrophysical applications resides in the fact that 3D structures of the cosmic web are identified starting from a point-like distribution, without making any assumption about its geometry or homogeneity. The 3D density field is built from the discrete distribution using the Delaunay Tessellation Field Estimator technique (DTFE; \citealp{Schaap2000, Cautun2011}), and it represents the input to the geometric 3D ridge extractor. {\sc DisPerSe} is a parameter-free and scale-free topologically motivated algorithm, based on discrete Morse and persistence theories. Voids, walls, and filaments are identified as individual components of the cosmic web and defined as distinct regions in the geometrical segmentation of space. The most significant structures can be selected according to their persistence ratio, which traces the significance of the topological connection between individual pairs of critical points and can be expressed in terms of number of standard deviations $\sigma$. 

For all the different approaches to select the cosmic web tracer populations (Section~\ref{Tracer populations: the combined redshift sample}), we adopt a fiducial persistence threshold of $N_{\sigma}=3$. We verify that our results remain robust against a more conservative threshold of $N_{\sigma}=5$, where main filaments persist and only small-scale features disappear. This is in agreement with the analysis on the sensitivity of the cosmic web reconstruction on the persistence threshold performed by \citet{Barsanti2026}. Despite the differences in the absolute number of filament segments ({\sc DisPerSE} arcs) and nodes, the association of the MAGPI galaxies within the large-scale environment and our results remain stable across the two thresholds.

Figure~\ref{CosmicWeb} shows the projected network of cosmic filaments (blue lines) for the G12, G15 and G23 regions at $0.20<z<0.42$ in comoving coordinates. Grey points are 86,587 galaxies from the combined redshift sample used to reconstruct the spine of the cosmic web, applying the mass-matched constant-number-density selection for the tracer population that was chosen as the fiducial approach. Red points are the 722 MAGPI galaxies with available stellar masses (Section~\ref{Selection criteria}), plotted on top of the reconstructed cosmic web. The G23 region is less densely sampled than G12 and G15 due to lower spectroscopic completeness (94.2\% for G23 compared to 98.5\% for G12 and G15 in the GAMA survey), fewer available stellar masses, and the absence of complementary DESI spectroscopy in this field. However, we find that excluding G23 does not affect our results. We therefore retain this region in our analysis to maximise statistical power.

\begin{figure}
\centering
\includegraphics[width=\columnwidth]{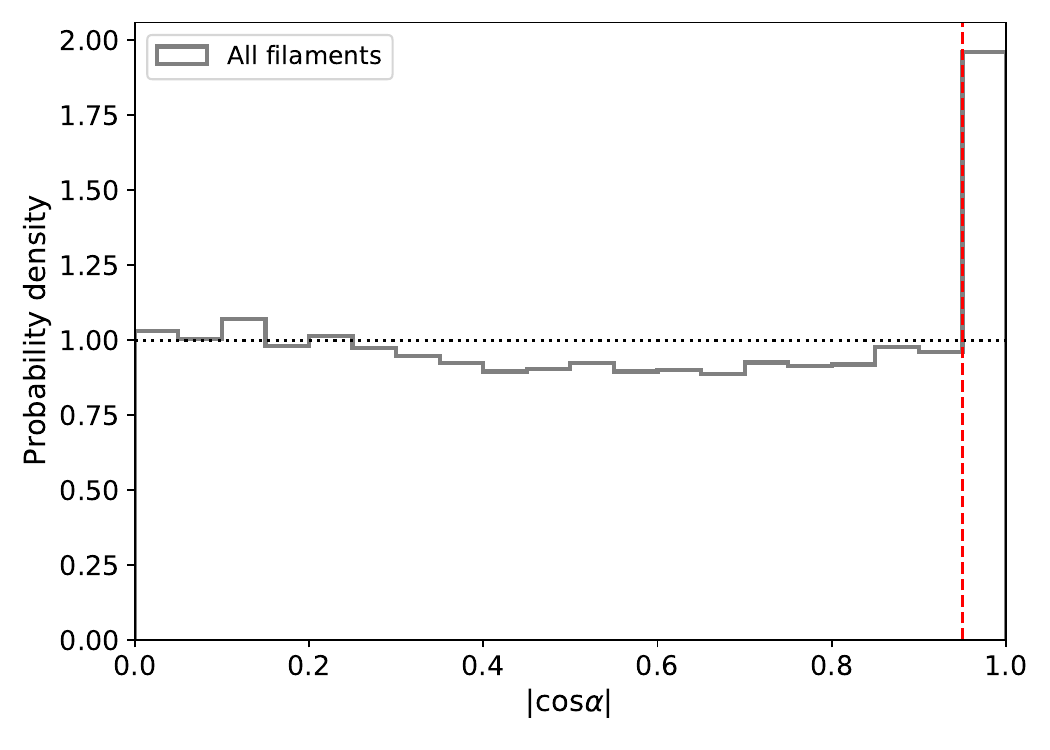}
\includegraphics[width=\columnwidth]{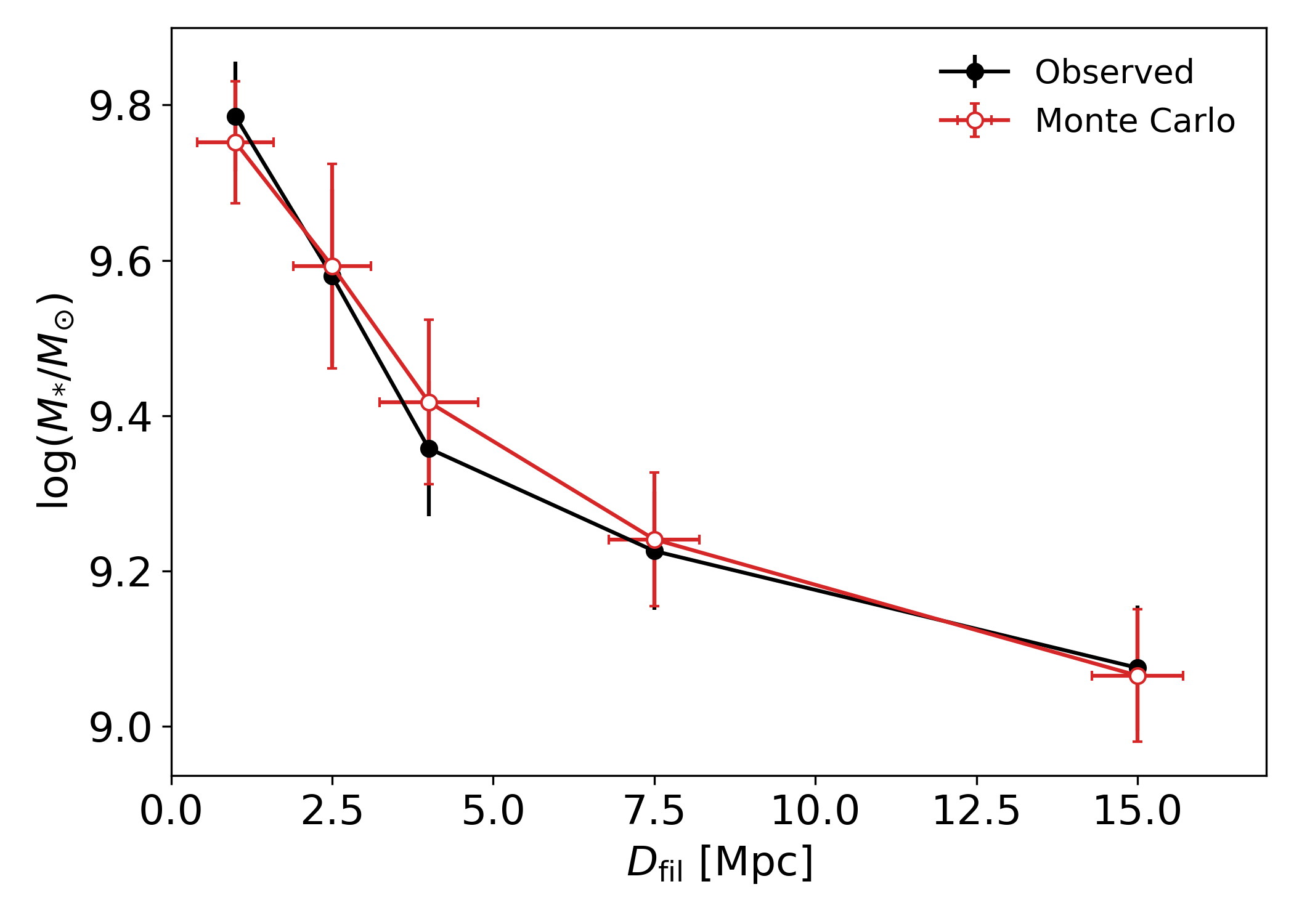}
\vspace*{-12pt}
\caption{Top panel: $|\cos\alpha|$ distribution for all filaments, where $\alpha$ is the angle between the filament and the line-of-sight. A uniform distribution is expected for isotropically oriented filaments, indicated by the black dotted line at a probability density of 1. An excess close to $|\cos\alpha|\sim1$ indicates a preferential alignment of filaments with the line-of-sight, arising from redshift-space distortions. The dashed red line marks $|\cos\alpha|=0.95$, above which MAGPI galaxies are re-assigned to filaments with $|\cos\alpha|<0.95$. Bottom panel: average stellar mass as a function of distance to the closest filament for 722 MAGPI galaxies at $0.20<z<0.42$ (black line). More massive galaxies are preferentially located closer to filaments. Error bars on stellar masses are 1$\sigma$ errors on the mean. Uncertainties in $D_{\rm fil}$ are estimated with a Monte Carlo approach that propagates the error in the line-of-sight galaxy positions due to peculiar velocities and in the filamentary network together (red line).}
\label{Filaments_plots}
\end{figure}

We assign each MAGPI galaxy to the closest filament using the smallest 3D Euclidean distance, $D_{\rm fil}$; if the projection point is beyond the start or end of the filament segment, we use the closest distance to the node. To correct for the FoG effect we apply a compression (Section~\ref{Tracer populations: the combined redshift sample}) before applying {\sc DisPerSe}, however this correction does not take into account boundary effects that can also lead to spurious filaments along the line-of-sight \citep{Welker2020} and has been found to perform poorly when applied to groups individually \citep{Kuchner2021}. Thus, to exclude possibly spurious filaments identified along the line-of-sight (see top panel of Figure~\ref{Filaments_plots}), we follow the same methodology as that used for the SAMI Galaxy Survey by \citet{Welker2020} and \citet{Barsanti2022}: MAGPI galaxies assigned to filaments with |$\cos\alpha$|\,>\,0.95 are reassigned to the closest filament with |$\cos\alpha$|\,<\,0.95, where $\alpha$ is the angle between the filament and the line-of-sight.

We recover the expected environmental trend that more massive galaxies are preferentially found closer to filament spines (bottom panel of Figure~\ref{Filaments_plots}), consistent with previous findings at low and intermediate redshifts \citep{Kraljic2018,Laigle2025}. A Spearman rank correlation test between stellar mass and filament distance yields a statistically significant negative correlation, with $\rho = -0.262$ and $p_{\rm S} = 8.6 \times 10^{-13}$. We test whether the trend remains significant accounting for errors on distances to the closest filament. Uncertainty in $D_{\rm fil}$ is estimated with a Monte Carlo approach that propagates the uncertainty in the line-of-sight galaxy positions due to peculiar velocities and in the filamentary network together. In each of 100 realisations, we perturb the line-of-sight comoving distance of every fiducial tracer with a Gaussian of width $\sigma_r = \sigma_v(1+z)/H(z)$, following \citet{Bel2014}. We adopt $\sigma_v = 250$\,km/s, which is within the range of 200–300 km s$^{-1}$ found for galaxies around filaments in simulations \citep{Kuchner2021}. We re-run Disperse on each catalogue of perturbed fiducial tracers. For each realisation, we recompute $D_{\rm fil}$ for every galaxy relative to the corresponding skeleton and recompute the mean stellar mass in each bin. We define the error on $D_{\rm fil}$ as the standard deviation of each bin mean across the realisations. The typical per-galaxy uncertainty is 0.5\,Mpc and $\sim15$\% of galaxies change bin in a typical realisation. The bin means shift by at most 0.7 times the original error bar, and the Spearman correlation changes from $\rho = -0.262$ to $-0.258 \pm 0.009$, so the trend is robust to uncertainties on $D_{\rm fil}$. 

\section{Methods}
\label{Methods}

\subsection{Orientation of the spin axes}
\label{Orientation of the spin axes}

To identify the orientation of the spin axes of galaxies, we follow the 3D thin-disc approximation implemented by \citet{Lee2007} and used in \citet{Kraljic2021} (see Section~2.6 of \citealp{Kraljic2021} for a summary of the formulae). This technique requires position angles and inclination angles. We use stellar and ionised gas kinematic PAs separately as described in Section~\ref{Kinematic position angles}. The galaxy axial ratio is measured from {\sc ProFound} in the $i$-band (Section~\ref{The MAGPI Survey}). The inclination angle is computed according to \citet{Haynes1984}, where for the intrinsic flatness parameter we use the value $q_0=0.2$, appropriate for disk-dominated systems \citep{Lambas1992,Rodriguez2013,Weijmans2014}. This assumption is justified by our selection of galaxies with well-resolved velocity fields and reliable kinematic position angles, which preferentially selects rotation-supported galaxies. We verified that our results are not sensitive to this choice by varying $q_0$ within the range $0.15$--$0.25$. The inclination angle is set to 90\degree\ if the axial ratio is lower than the intrinsic flatness parameter. The sign of the inclination angle is ambiguous, since it is not possible to determine whether the spin axis is pointing in projection towards or away from us. Following \citet{Kraljic2021}, we chose to consider only the positive sign for the cosine of the inclination angle. Our conclusions do not change if we take into account the two-fold ambiguity as proposed by \citet{Lee2011}, by assigning to each galaxy both signs (\citealp{Tempel2015}, appendix~A of \citealp{Kraljic2021}).

\subsection{Galaxy spin-filament angle}
\label{Galaxy spin-filament angle}

The galaxy spin--filament alignment is parametrised as the absolute value of the cosine of the angle between the galaxy spin axis and the closest filament in 3D Cartesian coordinates (e.g., \citealp{Tempel2013b,Kraljic2021}), given by
\begin{equation}
|\cos\gamma|=\frac{|\mathbf{L} \cdot \mathbf{r}|}{|\mathbf{L}| \cdot |\mathbf{r}|}
\end{equation}
where \textbf{L} is the galaxy spin axis identified as described in Section~\ref{Orientation of the spin axes} and \textbf{r} is the orientation vector of the filament. The quantity |$\cos\gamma$| lies in the range [0,1], with |$\cos\gamma$|=1 meaning the galaxy spin axis is parallel to the filament while |$\cos\gamma$|=0 means the galaxy spin axis is perpendicular to the filament. Schematic views of the cosmic web metrics and the associated vectors are reported in Figure~4 of \citet{Kraljic2018}, Figure~1 of \citet{Winkel2021}, and Figure~5 of \citet{Barsanti2022}.  

\section{Results}
\label{Results}

\subsection{Trends with $M_{\star}$}
\label{Trends with stellar mass}

We investigate the stellar mass dependence of galaxy spin--filament alignments at $z \sim 0.3$, focussing on the transition from parallel to perpendicular alignment with increasing stellar mass. We aim to quantify the transition stellar mass where the preferred spin orientation relative to large-scale filaments changes preference. We compare our results with the SAMI galaxy spin--filament alignments at $z\sim0.05$ to explore possible evolutionary trends in the signal over $\sim3$\,Gyr. 

\begin{figure*}
\centering
\vspace*{-18pt}
\includegraphics[width=14cm]{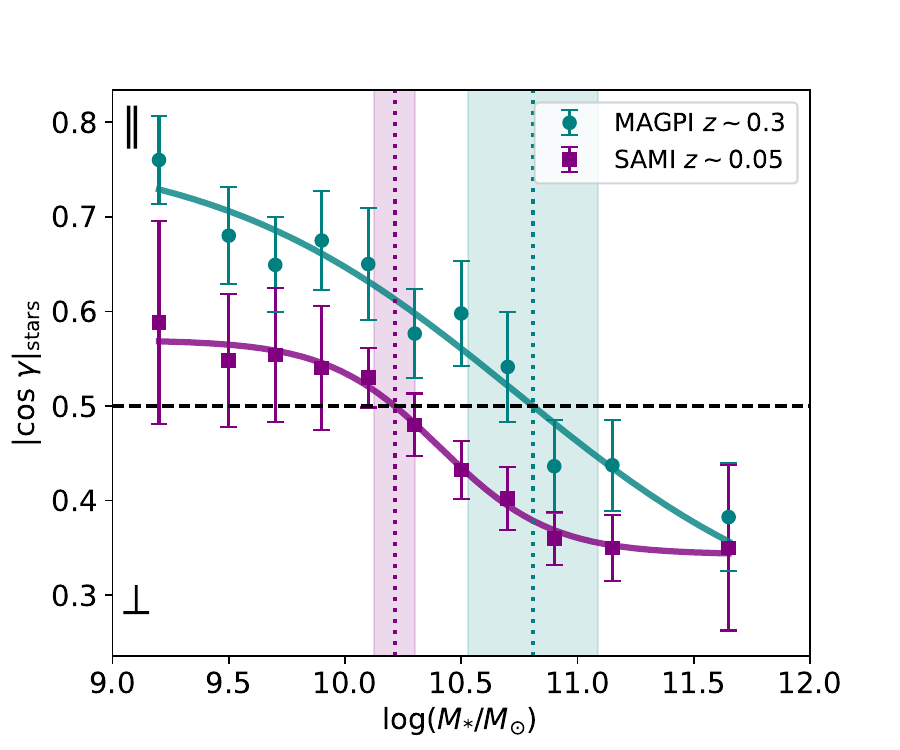}
\caption{Average |$\cos\gamma$| values as a function of bins in stellar mass for 342 MAGPI galaxies (teal circles) and 1388 SAMI galaxies (purple squares). The black dashed line marks $|\cos\gamma|=0.5$, the random orientation with respect to filaments. Standard errors on the mean are shown. The solid lines are the fitted sigmoid functions. The teal and purple dotted lines mark the transition stellar masses $\log(M_{\star,t}/M_{\odot})_{\rm MAGPI}$ and $\log(M_{\star,t}/M_{\odot})_{\rm SAMI}$ with 1$\sigma$ uncertainties shown as shaded areas. }
\label{cosgamma_mass}
\end{figure*}

Figure~\ref{cosgamma_mass} shows the average values of $|\cos\gamma|_{\rm stars}$ as a function of bins in stellar mass for MAGPI galaxies (selected in Section~\ref{Selection criteria}) and SAMI galaxies (selected in Section~\ref{SAMI galaxy survey}). At both $z\sim0.3$ and $z\sim0.05$ $|\cos\gamma|_{\rm stars}$ decreases monotonically with stellar mass. Applying the Spearman rank correlation test to the whole samples, we find significant correlations between $|\cos\gamma|_{\rm stars}$ and $M_{\star}$ for MAGPI ($\rho=-0.30$; $p_{\rm S}=0.015$) and SAMI ($\rho=-0.14$; $p_{\rm S}=0.001$). Although MAGPI has the larger correlation coefficient, its smaller sample results in a lower statistical significance than that of SAMI. However the trend is the same for MAGPI and SAMI: low-mass galaxies tend to have more parallel alignments ($|\cos\gamma|_{\rm stars}>0.5$), while high-mass galaxies have preferentially perpendicular alignments ($|\cos\gamma|_{\rm stars}<0.5$). MAGPI measurements of $|\cos\gamma|_{\rm stars}$ are systematically offset above SAMI across the full stellar mass range, indicating a tendency toward more parallel alignment at $z \sim 0.3$ compared to $z$\,$\sim$\,0 at fixed stellar mass.

In order to identify the transition mass we fit separate sigmoid functions to the MAGPI and SAMI mass-binned data. The mean trend of $|\cos\gamma|_{\rm stars}$ vs. $M_\star$ is well described by
\begin{equation}
|\cos\gamma|_{\rm stars}(M_\star)
=
A + \frac{B - A}{1 + \exp\left(\frac{\log M_\star - \log M_{\star,t}}{w}\right)}
\end{equation}
where $A$ and $B$ are the asymptotic values of $|\cos\gamma|_{\rm stars}$ at the high-mass and low-mass regimes, respectively. The parameter $w$ is the transition width which controls the sharpness of the transition in $\log M_\star$ space. We define the transition stellar mass, $\log M_{\star,t}$, as the stellar mass at which the fitted sigmoid crosses $|\cos\gamma|_{\rm stars}=0.5$, corresponding to the change from preferentially parallel ($|\cos\gamma|_{\rm stars}>0.5$) to preferentially perpendicular ($|\cos\gamma|_{\rm stars}<0.5$) alignments. The transition mass is therefore obtained by solving the fitted sigmoid for $|\cos\gamma|_{\rm stars}=0.5$. 

To estimate the uncertainty on the transition stellar mass, we performed a Monte Carlo sampling of the fitted sigmoid parameters. For each survey, $10^4$ sets of parameters $(A,B,\log M_{\star,\rm mid},w)$ were randomly drawn from the multivariate normal distribution defined by the best-fitting parameters and their covariance matrix returned by the sigmoid fit. For each realisation, we estimate $\log M_{\star,t}$. The uncertainty on the transition stellar mass was then taken as the standard deviation of the resulting distribution of transition masses, corresponding to a $1\sigma$ uncertainty. 

We find a transition mass for the MAGPI sample of 
$\log(M_{\star,t}/M_{\odot})_{\rm MAGPI} = 10.81 \pm 0.28$, and for the SAMI sample of 
$\log(M_{\star,t}/M_{\odot})_{\rm SAMI} = 10.22 \pm 0.09$. The difference between the two measurements is $\Delta \log(M_{\star,t}) = 0.6~\mathrm{dex}$. The uncertainty on this difference, computed by adding the individual errors in quadrature, is $\sigma_{\Delta} \approx 0.29$, corresponding to a significance of $\Delta \log(M_{\star,t}) / \sigma_{\Delta} \approx 2$. Thus, within MAGPI the transition from parallel to perpendicular spin–filament alignment occurs at a stellar mass $\sim 4$ times higher than in SAMI, providing marginal evidence for a shift in the transition mass between the two samples.

To assess the robustness of the difference in transition stellar mass between the MAGPI and SAMI samples, we performed a bootstrap analysis of the $|\cos\gamma|_{\rm stars}$ measurements. For each bootstrap realisation, we fitted the sigmoid function. Bootstrap realisations for which the sigmoid fit did not yield a valid crossing were discarded. We repeated this procedure $10^4$ times and compared the resulting transition masses by calculating $\Delta\log(M_{\star,t}) = \log(M_{\star,t})_{\rm MAGPI}-\log(M_{\star,t})_{\rm SAMI}$ for each realisation. Of the 9,863 valid bootstrap realisations, 99.5\% yielded a positive $\Delta\log(M_{\star,t})$, while only 0.5\% yielded a negative difference, strengthening our result that the transition mass is higher in MAGPI than in SAMI. 

\subsection{Mass-matched galaxy spin--filament alignments}
\label{Mass-matched spin flips}

\begin{figure}
\centering
\vspace*{-12pt}
\includegraphics[width=\columnwidth]{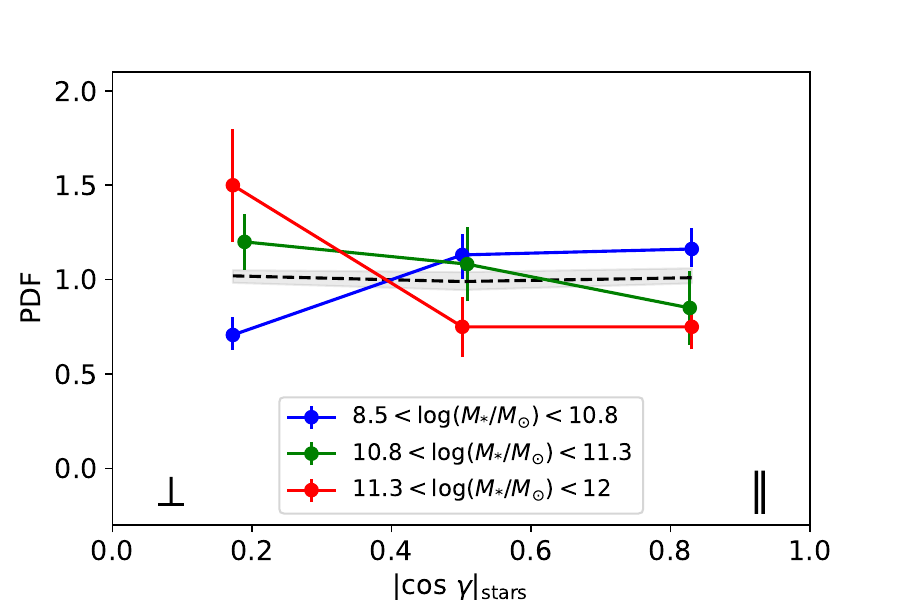}
\vspace*{-6pt}
\caption{PDF of stellar spin--filament alignments for 342 MAGPI galaxies divided into three stellar mass ranges: below, around, and above the inferred transition mass $\log(M_{\star,t}/M_{\odot})_{\rm MAGPI} = 10.81 \pm 0.28$. We show bootstrap error bars from 1000 random samples. The black dotted line and the shaded region represent the constructed null hypothesis with its 95\% confidence intervals. The PDF is skewed towards more perpendicular spin--filament alignments (i.e.\ lower values of $|\cos\gamma|_{\rm stars}$) for more massive galaxies and towards more parallel spin--filament alignments (i.e.\ higher values of $|\cos\gamma|_{\rm stars}$) for less massive galaxies.}
\label{fig:magpi_PDF}
\end{figure}

Following the analysis of \citet{Welker2020} and \citet{Barsanti2022} for SAMI galaxies, we investigate in Figure~\ref{fig:magpi_PDF} the probability distribution function (PDF) for $|\cos\gamma|_{\rm stars}$ of the 342 MAGPI galaxies grouped into three stellar mass ranges: $8.5<\log{(M_\star/M_{\odot})}<10.8$, $10.8<\log{(M_\star/M_{\odot})}<11.3$, and $11.3<\log{(M_\star/M_{\odot})}<12$, corresponding to galaxies below, around, and above the inferred transition mass $\log(M_{\star,t}/M_{\odot})_{\rm MAGPI} = 10.81 \pm 0.28$. The PDFs are normalised such that the mean value over the bins is unity. The error bars are estimated from the bootstrap method using 1000 sample realizations. To assess the statistical significance of each trend, we apply the one-sample Kolmogorov-Smirnov test (K-S test; \citealp{Lederman1984}) to test the null hypothesis that $|\cos\gamma|_{\rm stars}$ has a uniform distribution. In order to account for possible observational bias, we build the null hypothesis by generating 3000 randomised samples where the galaxy spins are fixed, but the galaxy positions (and thus the identification of their closest filaments) are shuffled \citep{Tempel2013a,Kraljic2021,Barsanti2022}. The median from the 3000 random samples is used as null hypothesis; the distributions constructed this way are nearly uniform, as expected. We take $p$-values ($p_{\rm KS}$) smaller than 0.05 to indicate that the distribution is significantly different from the null hypothesis. Table~\ref{KSresults} lists the number of galaxies for each stellar mass range, the mean $|\cos\gamma|_{\rm stars}$ value, and the $p$-value from the K-S test. 

We find that the PDF is significantly skewed towards more perpendicular spin--filament alignments (i.e.\ lower values of $|\cos\gamma|_{\rm stars}$) for more massive galaxies ($\log{(M_\star/M_{\odot)}}>10.8$) and towards more parallel spin--filament alignments (i.e.\ higher values of $|\cos\gamma|_{\rm stars}$) for less massive galaxies. Using the two-sample K-S test, we check the null hypothesis that the $|\cos\gamma|_{\rm stars}$ distributions are drawn from the same parent population. This test yields $p_{\rm 2 K-S}<0.05$ (Table~\ref{KSresults}) when comparing the lowest-mass bin with both more massive populations, indicating that their $|\cos\gamma|_{\rm stars}$ distributions are unlikely to be drawn from the same parent population. In contrast, the distributions for galaxies with $10.8<\log(M_{\star}/M_\odot)<11.3$ and $11.3<\log(M_{\star}/M_\odot)<12.0$ are statistically consistent with belonging to the same parent population.

\begin{figure}
\centering
\includegraphics[width=\columnwidth]{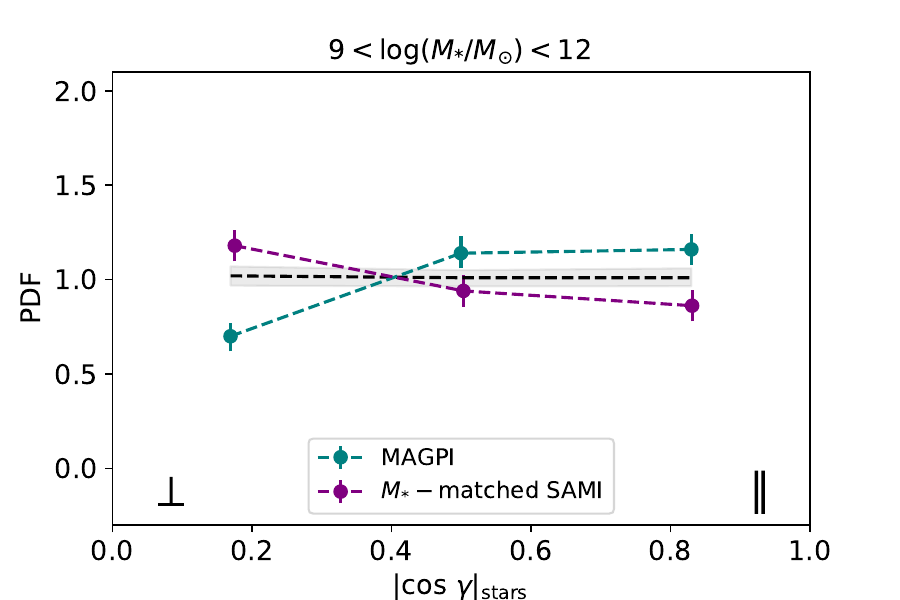}
\vspace*{-6pt}
\caption{PDFs of stellar spin--filament alignments for 300 MAGPI galaxies and stellar-mass-matched 300 SAMI galaxies. The MAGPI galaxies exhibit a $|\cos\gamma|_{\rm stars}$ distribution skewed towards more parallel spin--filament alignments, whereas the $M_{\star}$-matched SAMI galaxies show a tendency towards more perpendicular alignments.}
\label{fig:magpi_comparison}
\end{figure}

We compare the spin--filament alignment signal in the MAGPI Survey with that measured in the SAMI Galaxy Survey in Figure~\ref{fig:magpi_comparison} for galaxies with $9<\log(M_{\star}/M_\odot)<12$. We construct a stellar-mass-matched SAMI sample following a procedure similar to that adopted by \cite{Zovaro2024} and \cite{Barsanti2025a}. For each MAGPI galaxy, we identify the unique SAMI galaxy with the closest stellar mass. This results in matched samples comprising 300 MAGPI galaxies and 300 stellar-mass-matched SAMI galaxies. A K-S test confirms that the stellar mass distributions of the two samples are statistically indistinguishable. The MAGPI galaxies exhibit a $|\cos\gamma|_{\rm stars}$ distribution skewed towards more parallel spin--filament alignments, whereas the $M_{\star}$-matched SAMI galaxies show a tendency towards more perpendicular alignments (Table~\ref{KSresults}). A two-sample K-S test indicates that the two distributions are significantly different ($p_{\rm 2KS}=0.007$). This suggests that, at fixed stellar mass, galaxies at $z\sim0.3$ are more likely to exhibit parallel spin--filament alignments than their counterparts at $z\sim0.05$, in agreement with the MAGPI vs SAMI comparison shown in Figure~\ref{cosgamma_mass}.

\begin{table*}
\centering
\caption{K-S test results for galaxy spin--filament alignments. Column~1 lists the galaxy survey, column~2 the stellar mass ranges, column~3 the number of galaxies in each range, column~4 the spin tracer, column~5 the mean value $\langle |\cos\gamma| \rangle$, column~6 the $p$-value from the K--S test, and column~7 the alignment tendency where a significant deviation from the reconstructed uniform distribution is detected. Columns~8 and~9 give the samples compared using a two-sample K--S test, with the corresponding $p$-value listed in column~10. Significant $p$-values ($p<0.05$) are highlighted in bold.}
\label{KSresults}

\begin{tabular}{lcccccc|ccc}
\toprule Galaxy survey & $\log(M_\star/M_\odot)$ & $N_{\rm gal}$ & Spin type & $\langle |\cos\gamma| \rangle$ & $p_{\rm KS}$ & Alignment & Sample 1 & Sample 2 & $p_{\rm 2KS}$ \\ 
\midrule
MAGPI & $[8.5,10.8]$  & 224 & stars & $0.563\pm0.020$ & \textbf{0.009} & $\parallel$ & $[8.5,10.8]$  & $[10.8,11.3]$ & \textbf{0.001}\\
MAGPI & $[10.8,11.3]$ & 94  & stars & $0.441\pm0.031$ & \textbf{0.038} & $\perp$     & $[10.8,11.3]$ & $[11.3,12.0]$ & 0.315 \\
MAGPI & $[11.3,12.0]$ & 24  & stars & $0.427\pm0.110$ & \textbf{0.045} & $\perp$     & $[11.3,12.0]$ & $[8.5,10.8]$  & \textbf{0.007} \\
\midrule
MAGPI & $[9.0,12.0]$  & 300 & stars & $0.573\pm0.017$ & $\mathbf{4\times10^{-4}}$ & $\parallel$ & MAGPI  & $M_{\star}$--matched SAMI & \textbf{0.007}\\
$M_{\star}$--matched SAMI & $[9.0,12.0]$  & 300 & stars & $0.447\pm0.019$ & \textbf{0.018} & $\perp$ &  &  & \\
\midrule
MAGPI & $[7.0,10.8]$  & 453 & gas & $0.608\pm0.013$ & \textbf{0.003} & $\parallel$ & $[7.0,10.8]$  & $[10.8,11.3]$ & \textbf{0.005}\\
MAGPI & $[10.8,11.3]$ & 51 & gas & $0.481\pm0.038$ & \textbf{0.031} & $\perp$     & $[10.8,11.3]$ & $[11.3,12.0]$ & 0.749 \\
MAGPI & $[11.3,12.0]$ & 19  & gas & $0.460\pm0.062$ & \textbf{0.046} & $\perp$     & $[11.3,12.0]$ & $[7.0,10.8]$  & \textbf{0.006} \\
\bottomrule
\end{tabular}
\end{table*}

\subsection{Stellar spin--filament vs ionised gas spin--filament alignments}
\label{Stellar vs gas}

\begin{figure}
\centering
\vspace*{-12pt}
\includegraphics[width=\columnwidth]{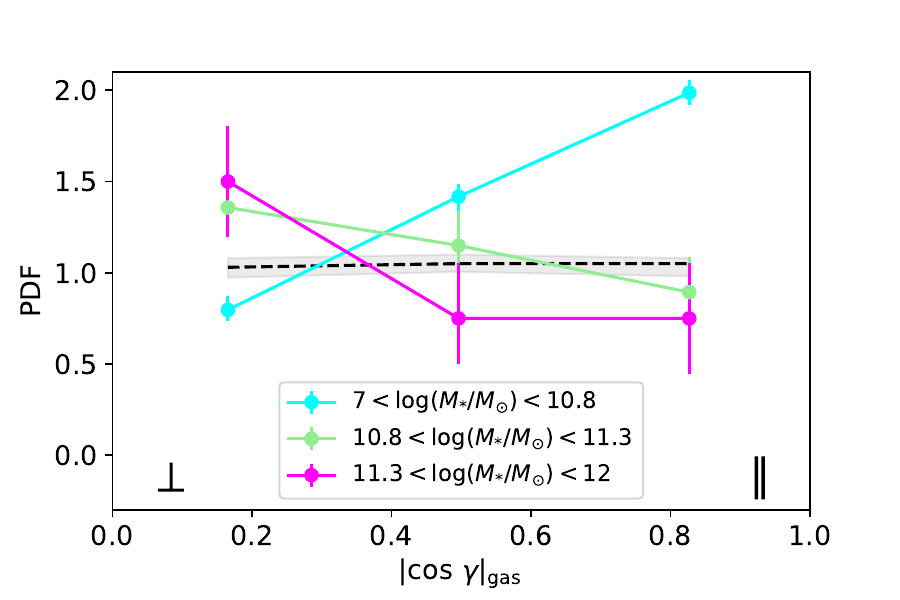}
\vspace*{-12pt}
\caption{PDFs of ionised gas spin--filament alignments for 523 MAGPI galaxies divided in stellar mass. Galaxies with $7<\log(M_\star/M_{\odot})<10.8$ show a tendency to more parallel alignments, while more massive galaxies show more perpendicular trends.}
\label{fig:magpi_gas}
\end{figure}

We explore the orientation of ionised gas spin axes (measured in Section~\ref{Ionised gas kinematics}) with respect to the closest filament, allowing us to probe the very low-mass tail of spin--filament alignments for MAGPI galaxies with $7<\log(M_\star/M_{\odot})<8.5$. Figure~\ref{fig:magpi_gas} shows the PDFs of the ionised gas spin--filament alignments for 523 MAGPI galaxies divided into three $M_\star$ ranges: below, around, and above the inferred transition mass $\log(M_{\star,t}/M_{\odot})_{\rm MAGPI} = 10.81 \pm 0.28$. The results from the K-S tests are listed in Table~\ref{KSresults}. Galaxies with $7<\log(M_\star/M_{\odot})<10.8$ tend to show a significant parallel tendency, while more massive galaxies have more perpendicular alignments. The $|\cos\gamma|_{\rm gas}$ distribution for $7<\log(M_\star/M_{\odot})<10.8$ is significantly different to the one for $10.8<\log(M_\star/M_{\odot})<11.3$ ($p_{\rm 2\,K-S}=0.005$) and to the one for $11.3<\log(M_\star/M_{\odot})<12$ ($p_{\rm 2\,K-S}=0.006$), while these latter two are consistent with being drawn from the same parent population. These findings are in agreement with those obtained in Figure~\ref{fig:magpi_PDF} for stellar spin--filament alignments. This is expected, since 242 MAGPI galaxies have reliable measurements of both ${\rm PA}_{\rm stars}$ and ${\rm PA}_{\rm gas}$, and $\sim60\%$ of these satisfy $\Delta {\rm PA} = |{\rm PA}_{\rm stars}-{\rm PA}_{\rm gas}| < 15^\circ$, indicating a substantial degree of kinematic alignment between the stellar and gaseous components.

For 242 MAGPI galaxies with measured stellar spin--filament and ionised gas spin--filament alignments, we compare whether gas or stellar spin align more strongly with the closest filament. We define the offset
\begin{equation}
    \Delta |\cos\gamma| = |\cos\gamma|_{\rm gas} - |\cos\gamma|_{\rm stars},
\end{equation}
where $\Delta |\cos\gamma| > 0$ indicates that the gas spin is more strongly aligned with the filament than the stellar spin, while $\Delta |\cos\gamma| < 0$ indicates the opposite. This differential quantity probes the relative coupling of gas and stars to the surrounding large-scale structure. MAGPI galaxies exhibit a positive mean offset ($+$0.058) and a positive median offset ($+$0.097), indicating that, on average, the gas component is more strongly aligned with the cosmic web than the stellar component. In contrast, if we compare MAGPI with 242 stellar-mass-matched SAMI galaxies (Table~\ref{StellarVSgas_results}), the SAMI sample shows a slightly negative mean offset ($-$0.046) and a median offset consistent with zero, suggesting a reduced gas--filament alignment coupling advantage at low redshift and a more consistent coupling between gas and stellar angular momentum with respect to the surrounding large-scale structure.

\begin{figure}
\centering
\includegraphics[width=\columnwidth]{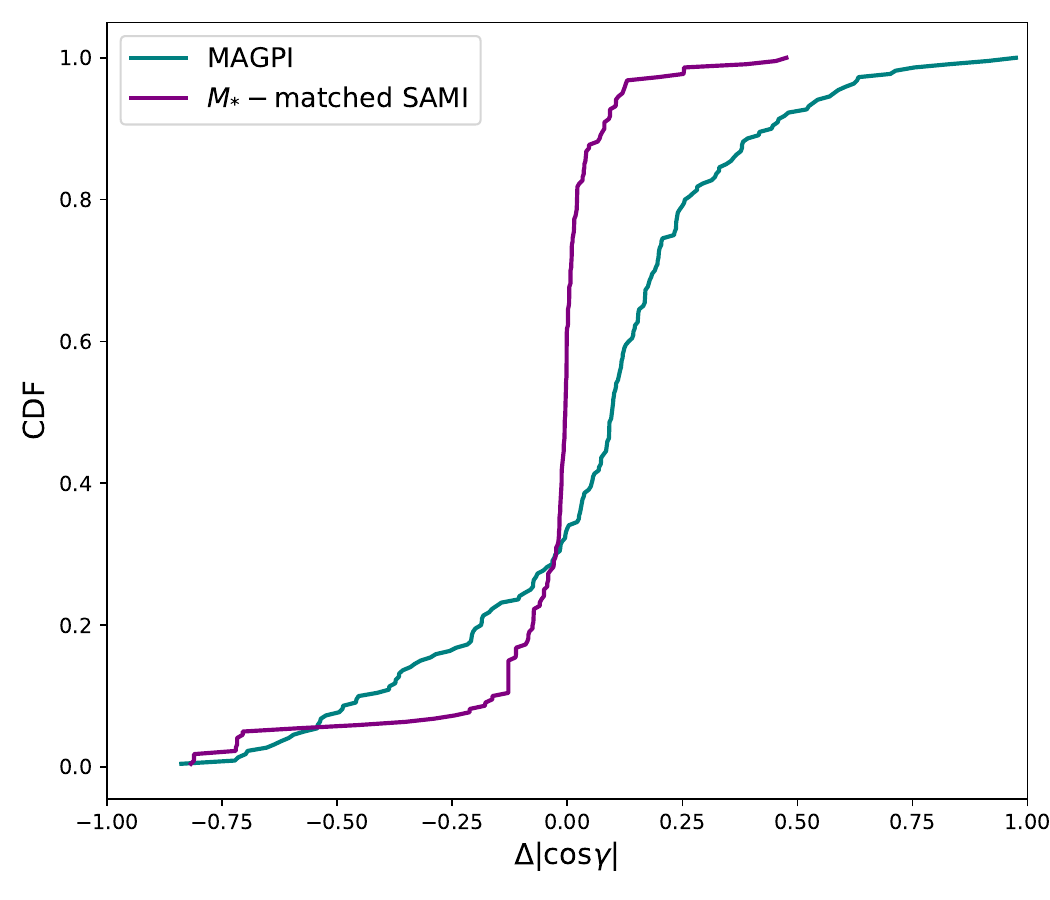}
\vspace*{-12pt}
\caption{CDFs for $\Delta|\rm cos\gamma|=|\cos\gamma|_{\rm gas} -|\cos\gamma|_{\rm stars}$ distributions for 242 MAGPI galaxies compared to 242 stellar-mass-matched SAMI galaxies. The distributions are
significantly different according to the two-sample K-S test, with MAGPI galaxies showing a systematic excess of positive $\Delta|\rm cos\gamma|$ values, indicating stronger gas–filament alignment than stars relative to the SAMI sample.}
\label{CDF_magpi_sami}
\end{figure}

The difference between the MAGPI and SAMI mean offsets (0.104) is significant at approximately $4\sigma$. Figure~\ref{CDF_magpi_sami} shows the cumulative distribution functions (CDFs) of $\Delta |\cos\gamma|$ for MAGPI and the mass-matched SAMI sample. A two-sample K-S test confirms that the full distributions are significantly different (Table~\ref{StellarVSgas_resultsKS}).

The larger standard deviation in the MAGPI sample (see Table~\ref{StellarVSgas_results}) may reflect a broader diversity of dynamical states at $z \sim 0.3$, although part of this increased scatter may also arise from the larger uncertainties associated with measuring kinematic PAs at higher redshift. The  narrower distribution of the SAMI sample is consistent with more dynamically evolved systems at low redshift and smaller uncertainties in kinematic PAs, where the relative gas--stellar alignment with the closest filament is weaker and more homogenised.

\begin{table*}
\centering
\caption{Stellar spin--filament vs ionised gas spin--filament alignments. Column~1 lists the galaxy survey, column~2 the total number of galaxies, columns~3,~4 and~5 report the mean, median and standard deviation of the $\Delta|\rm cos\gamma|$ distributions. Columns~6 and~7 report the number of galaxies with $\Delta|\rm cos\gamma|>0$ and $\Delta|\rm cos \gamma|<0$.}
\label{StellarVSgas_results}
\begin{tabular}{lcccccc}
\toprule Galaxy survey & $N_{\rm gal}$ &  $\langle \Delta|\cos\gamma| \rangle$ & $\widetilde{\Delta |\cos\gamma|}$ & $\sigma({\Delta |\cos\gamma|})$ & $N_{\rm \Delta|\cos\gamma|>0}$ (\%) & $N_{\rm \Delta|\cos\gamma|<0}$ (\%)  \\ 
\midrule
MAGPI &  242 & $+0.058$ & $+0.097$ & 0.323 & 157 (64.9\%) & 85 (35.1\%) \\
$M_{\star}$--matched SAMI &  242  & $-0.046$ & $-0.004$ & 0.193 & 95 (39.3\%)     & 147 (60.7\%)  \\
\bottomrule
\end{tabular}
\end{table*}

\begin{table}
\centering
\caption{Two-sample K--S test for stellar spin--filament vs ionised gas spin--filament alignments. Columns~1 and~2 give the comparing samples with the number of galaxies in brackets. The corresponding $p$-value is listed in column~3. Significant $p$-values ($p<0.05$) are in bold.}
\label{StellarVSgas_resultsKS}
\begin{tabular}{ccc}
\toprule Sample 1 & Sample 2 & $p_{\rm 2KS}$ \\ 
\midrule
MAGPI (242) & $M_{\star}$--matched SAMI (242) & $\mathbf{10^{-23}}$\\
\bottomrule
\end{tabular}
\end{table}

\section{Discussion}
\label{Discussion}

Our results suggest evolution in galaxy spin--filament alignments over the last $\sim3$\,Gyr, reflected in a higher transition stellar mass (by $\sim0.6$ dex at $z\sim0.3$ relative to $z\sim0.05$), a shift in the global alignment signal from preferentially parallel to more perpendicular orientations, and a stronger gas--filament coupling at $z\sim0.3$ with respect to $z\sim0.05$, where the ionised gas component is more closely aligned with the filamentary network than the stellar component. A possible interpretation is that galaxies at $z\sim0.3$ remain more strongly coupled to the anisotropic matter flows associated with the cosmic web. Cosmological simulations show that gas can accrete through  cold coherent inflows along cosmic filaments rather than through isotropic infall, delivering angular momentum to galaxies and contributing to disc growth \citep{Kerevs2005,Dekel2006,Kerevs2009,Pichon2011}. In this regime, ongoing filamentary accretion can continuously supply gas whose angular momentum is correlated with the surrounding large-scale structure. This may enhance gas--filament alignments and help to maintain preferentially parallel spin orientations to higher stellar masses, consistent with the role of large-scale tidal torques in shaping galaxy angular momentum \citep{Pichon2011}. As galaxies evolve toward lower redshift, repeated merger events and secular processes progressively perturb the angular momentum initially acquired through filamentary inflows. These mechanisms can reorient galaxy spins at lower stellar masses, promoting the emergence of perpendicular alignments and reducing the differential alignment between the gaseous and stellar components with respect to the closest filament. This trend is consistent with a gradual weakening of the coupling between galaxy angular momentum and the cosmic web, driven by the cumulative effects of past accretion, isotropic infall, mergers and dynamical evolution.

Observational studies of HI also provide evidence for connections between gas kinematics and the cosmic web. \citet{BlueBird2020} found a preferential alignment of HI spins with nearby filaments, while \citet{Tudorache2022} reported that this alignment depends on the HI-to-stellar mass ratio, suggesting that gas accretion and merger history may influence the observed signal. These results suggest that the influence of filamentary accretion may depend on the gas phase \citep{Jung2025}.

\citet{Codis2012} provide the foundational picture in which dark matter halo spin--filament alignments depend on mass and redshift. In this framework, low-mass systems acquire angular momentum through coherent anisotropic accretion towards filaments, leading to parallel alignments, while high-mass systems grow predominantly through mergers preferentially along filament axes, converting orbital angular momentum into spin and driving perpendicular alignments. This model predicts a transition mass that evolves with redshift, $M_{t}(z) \propto (1+z)^{-\gamma}$, with $\gamma \sim 2.5$. Our finding of a higher transition stellar mass in MAGPI relative to SAMI is qualitatively consistent with this prediction, with the difference marginally significant at the $\sim2\sigma$ level. Within this framework, the stronger parallel alignments and enhanced gas--filament coupling observed in MAGPI galaxies may reflect the cumulative imprint of coherent filamentary accretion at earlier cosmic times, rather than a higher instantaneous accretion rate at $z\sim0.3$. Galaxies observed at $z\sim0.3$ may retain more information about the preferred accretion directions established at $z\gtrsim1.5$ during the epoch of more efficient coherent accretion, whereas subsequent merger-driven reorientation and dynamical evolution becomes increasingly important toward $z$\,$\sim$\,0.

An additional interpretation is provided by \citet{Tobar2026}, who demonstrate that dark matter halo spin--filament alignments exhibit a similar mass dependence but show no significant evolution between $z=1$ and $z=0$. In their picture, major mergers are the primary mechanism capable of significantly reorienting halo spins, driving high-mass systems toward perpendicular configurations. The stability of the halo alignment signal contrasts with the evolution observed in our galaxy sample, including a higher transition mass and stronger parallel alignments in MAGPI relative to SAMI. This suggests that the observed evolution may arise primarily at the baryonic level, where stellar and gaseous angular momentum remain sensitive to ongoing gas accretion, mergers, flybys, feedback, and internal dynamical processes, even if the underlying halo orientation is relatively stable, reflecting different physical processes governing dark matter and baryonic angular momentum. Moreover, the measurements of halo spins are difficult to constrain \citep{Contreras2017} and can exhibit significant radial variations, with the angular momentum of the inner regions, where galaxies reside, differing from that of the halo as a whole \citep{Bett2016}. Consequently, there can be galaxy--halo spin misalignments, especially for elliptical galaxies \citep{Teklu2015,Valenzuela2024}.

\citet{Jhee2026} further extend this picture by showing that halo spin evolution can be driven by coherent filament-induced torques even in the absence of major mergers. In their framework, halos experience distinct mass flow regimes within and around filaments that generate continuous, non-stochastic modifications of angular momentum. This is consistent with the enhanced gas--filament alignment observed in MAGPI: the ionised gas component is more responsive to ongoing anisotropic accretion and filamentary torques than the stellar component, which reflects the integrated assembly history. As a result, galaxies at $z\sim0.3$ retain a stronger coupling between angular momentum and filamentary inflow, whereas at $z$\,$\sim$\,0 this coupling is reduced, consistent with the weaker gas--stellar differential alignment observed in SAMI.

Together, these studies support an evolutionary picture in which galaxy spin--filament alignments are shaped by the interplay between (i)~coherent filamentary accretion, (ii)~merger-driven spin reorientation, and (iii)~continuous filamentary torques operating even in the absence of mergers. Our observations suggest that the balance between these processes evolves over cosmic time, with coherent accretion and filamentary torques playing a more prominent role at $z\sim0.3$, while merger-driven and cumulative evolutionary effects become increasingly dominant toward $z$\,$\sim$\,0. 

Finally, the spin--filament alignments recovered in this study are relatively weak preferences (even when statistically significant) and based on limited galaxy samples. We note that the observational study of spin--filament alignments is still a challenging field. At low redshift this is because of the intrinsic weakness of the signal \citep{Codis2018}, while at intermediate redshift it is due to the lack of large IFS datasets and of large spectroscopic redshift surveys to reconstruct the cosmic web. Thus, our work provides only hints for galaxy evolution mechanisms and future larger samples are required to confirm constraints across cosmic time for galaxy alignments. 

\section{Summary and conclusions}
\label{Summary and conclusions}

We present measurements of galaxy spin–filament alignments at 
$0.20<z<0.42$ using the MAGPI Survey, exploiting spatially-resolved stellar and ionised gas kinematics to reconstruct 3D spin axes. Cosmic
filaments are identified from a combined redshift sample incorporating the
GAMA, MAGPI Redshift Survey, and DESI datasets. The cosmic web is reconstructed using different selection approaches for defining the tracer population in the three GAMA regions G12, G15 and G23: full sample, fixed $M_{\star}$ cut, redshift-dependent $M_{\star}$ cut, weights, constant number density and $M_{\star}$--matched constant-number-density. The latter, which is characterised by 88,485 galaxies, is selected as the fiducial approach, since it minimises redshift-dependent selection effects and provides the best
compromise between sample homogeneity and spatial sampling. Stellar spin--filament alignments are available for 342 MAGPI galaxies at $0.20<z<0.42$ with $8.5<\log(M_\star/M_{\odot})<12$. Ionised gas spin--filament alignments are available for 523 MAGPI galaxies with $7<\log(M_\star/M_{\odot})<12$. There are 242 MAGPI galaxies with both measurements. Our main results are as follows:
\begin{enumerate}
    \item[(i)] We detect a flipping of stellar spin--filament alignments and ionised gas spin--filament alignments as a function of stellar mass at $z\sim0.3$: low-mass galaxies preferentially exhibiting spins parallel to their host filaments, while high-mass galaxies tend towards perpendicular orientations.  
    \item[(ii)] The transition mass for MAGPI galaxies is measured to be $\log(M_{\star,t}/M_{\odot})\simeq10.81$, approximately 0.6 dex higher than the value measured from the SAMI Galaxy Survey at $z\sim0.05$, with the difference being marginally significant at the $\sim2\sigma$ level. Bootstrap analysis gives 99.5\% confidence that the transition mass is higher for MAGPI than SAMI.
    \item[(iii)]  At fixed stellar mass, we find that MAGPI galaxies exhibit a systematically stronger parallel alignment relative to SAMI galaxies. Comparing the $|\cos\gamma|_{\rm stars}$ distribution for MAGPI galaxies with that of a $M_{\star}$--matched SAMI galaxy sample, we confirm that MAGPI galaxies tend to be more parallel aligned while SAMI galaxies are more perpendicularly aligned.
    \item[(iv)] Comparing stellar spin--filament alignments and ionised gas spin--filament alignments, we find that the gaseous component in MAGPI galaxies is more closely aligned with cosmic filaments than the stellar component relative to an $M_{\star}$--matched SAMI sample. On the other hand, for the SAMI sample the relative ionised gas--stellar alignment with the closest filament is more homogenised, with a slight preference for the stellar component to be more strongly aligned with the closest filament.
\end{enumerate}

These results provide hints for evolution in the transition stellar mass, the global alignment trend from more parallel to more perpendicular orientations, and the relative alignment between stellar and gaseous components with respect to the closest filament from $z\sim0.3$ to $z$\,$\sim$\,0. The observed trends are consistent with a scenario in which coherent gas accretion towards filaments plays a more important role at intermediate redshift, while the cumulative effects of mergers and secular evolution leads to a transition toward perpendicular alignments at lower stellar masses
and progressively weaken the direct imprint of the cosmic web on the gaseous component at later cosmic times.

%Our findings should be interpreted with caution. Although several of the observed trends are statistically significant, the relatively small number of galaxies with reliable spin measurements for the MAGPI Galaxy Survey ($\sim$400) and the combined inhomogeneous spectroscopic redshift measurements used to reconstruct the cosmic web limits the precision with which evolutionary scenarios can be constrained. Consequently, our results should be regarded only as suggestive hints for a possible evolution in spin--filament alignments from $z\sim0.3$ to $z$\,$\sim$\,0 rather than definitive proof of the underlying evolving physical mechanisms.

The next generation of high-multiplex single-fibre spectroscopic datasets and integral field spectroscopy surveys will dramatically improve evolutionary constraints in galaxy spin--filament alignments. At low redshift, the Hector Galaxy Survey \citep{Bryant2024,Oh2025} will observe IFS data for 15,000 galaxies, and will be combined with the vast upcoming 4MOST spectroscopic redshift surveys for millions of galaxies, such as the Wide Area VISTA Extra-galactic Survey (WAVES; \citealp{Driver2016,Kaur2025}) and the 4MOST Hemisphere Survey of the nearby Universe (4HS; \citealp{Taylor2023}) for mapping the large-scale structure at extremely high completeness. In addition, the 4MOST Spin Survey (Supplementary Programme 44; PI: Barsanti) will enable measurements of galaxy spin--filament alignments for substantially larger samples across the southern sky, by placing additional fibres on 4HS and WAVES galaxies. At higher redshift, the Wide-field Spectroscopic Telescope (WST; \citealp{Mainieri2024}) will extend such studies to $z\sim1$, allowing statistically robust measurements of spin--filament alignments and their evolution over the past $\sim8$\,Gyr. The future observational strategy proposed by \citet{Barsanti2025b} will combine large IFS samples with sub-Mpc cosmic web reconstructions, providing an unprecedented view of how galaxies acquire and evolve their angular momentum within the large-scale structure.

\section*{Acknowledgments}

We thank the anonymous referee, whose suggestions helped to clarify and improve the content of this work. We thank the ESO staff, and in particular the staff at Paranal Observatory, for carrying out the MAGPI observations. MAGPI targets were selected from GAMA. GAMA is a joint European-Australasian project based around a spectroscopic campaign using the Anglo-Australian Telescope. GAMA was funded by the STFC (UK), the ARC (Australia), the AAO, and the participating institutions. GAMA photometry is based on observations made with ESO Telescopes at the La Silla Paranal Observatory under programme ID 179.A-2004, ID 177.A-3016. 
This research was supported by the Australian Research Council Centre of Excellence for All Sky Astrophysics in 3 Dimensions (ASTRO~3D, CE170100013). The SAMI Galaxy Survey is based on observations made at the Anglo-Australian Telescope. The Sydney-AAO Multi-object Integral field spectrograph (SAMI) was developed jointly by the University of Sydney and the Australian Astronomical Observatory, and funded by ARC grants FF0776384 (Bland-Hawthorn) and LE130100198. The SAMI input catalogue is based on data taken from the Sloan Digital Sky Survey, the GAMA Survey, and the VST/ATLAS Survey. The SAMI Galaxy Survey website is http://sami-survey.org/. CF is the recipient of an Australian Research Council Future Fellowship (project number FT210100168) funded by the Australian Government.
CL, JTM and CF are the recipients of funding from the Australian Research Council (ARC) Discovery Project DP210101945. ET acknowledges funding from the HTM (grant TK202), ETAg (grant PRG3034) and the EU Horizon Europe (EXCOSM, grant No. 101159513). This study uses data provided by AAO Data Central (http://datacentral.org.au/) and the \href{http://www.python.org}{Python} programming language \citep{vanrossum1995}. We acknowledge the use of
{\sc \href{https://pypi.org/project/numpy/}{numpy}} \citep{harris+2020},
{\sc \href{https://pypi.org/project/scipy/}{scipy}} \citep{jones+2001},
{\sc \href{https://pypi.org/project/matplotlib/}{matplotlib}} \citep{hunter2007},
{\sc \href{https://pypi.org/project/astropy/}{astropy}} \citep{astropyco+2013}, {\sc \href{https://pypi.org/project/pyvista/}{pyvista}} \citep{Sullivan2019} and {\sc \href{http://www.star.bris.ac.uk/~mbt/topcat/}{topcat}} \citep{taylor2005}. We acknowledge the use of AI language models (OpenAI ChatGPT and Anthropic Claude) for some text refinement and coding debugging.

\section*{Data availability}
The MAGPI raw data (and a basic data reduction) are available through the \href{http://archive.eso.org/cms.html}{ESO Science Archive Facility}. MAGPI emission line data products are available at \href{https://docs.datacentral.org.au/magpi/}{MAGPI Data Central}. The SAMI reduced data underlying this article are publicly available at \href{https://docs.datacentral.org.au/sami}{SAMI Data Release 3} \citep{Croom2021}. Ancillary data comes from the \href{http://gama-survey.org}{GAMA Data Release 4} \citep{Driver2022}. The first data release (DR1) of the Dark Energy Spectroscopic Instrument (DESI) can be accessed through the DESI public data repository \href{ https://data.desi.lbl.gov/public/dr1}{https://data.desi.lbl.gov/public/dr1}.

\section*{Author Contribution Statement}

Stefania Barsanti devised the project and drafted the paper. AB measured MAGPI ionised gas mean velocity maps. SB measured stellar masses for MAGPI and for the MAGPI Redshift Survey. MSO contributed to the observations, reduced all data and measured all spectroscopic redshifts for the MAGPI Redshift Survey. MC, SMC and ET contributed to data analyses and interpretation of the results. CF contributed significantly to MAGPI data processing. KEH, CDPL, JTM and EW are the MAGPI Principal Investigators. All authors discussed the results and commented on the manuscript. 


\bibliographystyle{mnras}
\bibliography{biblioMAGPI} % if your bibtex file is called example.bib

%%%%%%%%%%%%%%%%%%%%%%%%%%%%%%%%%%%%%%%%%%%%%%%%%%

\section*{Affiliations}
\noindent
$^{1}${Sydney Institute for Astronomy (SIfA), School of Physics, The University of Sydney, NSW 2006, Australia}\\
$^{2}${International Centre for Radio Astronomy Research, University of Western Australia, 7 Fairway, Crawley, WA 6009, Australia}\\
$^{3}${Research School of Astronomy and Astrophysics, Australian National University, Cotter Road, Weston Creek, ACT 2611, Australia}\\
$^{4}${School of Physics, University of New South Wales, Sydney, NSW 2052, Australia}\\
$^{5}${Department of Physics, Institute for Computational Cosmology, Durham University, South Road, Durham DH13LE, United Kingdom}\\
$^{6}${Department of Physics, Centre for Extragalactic Astronomy, Durham University, South Road, Durham DH13LE, United Kingdom}\\
$^{7}${School of Mathematical and Physical Sciences, Macquarie University, Sydney, NSW 2109, Australia}\\
$^{8}${Astrophysics and Space Technologies Research Centre, Macquarie University, Sydney, NSW 2109, Australia}\\
$^{9}${Tartu Observatory, University of Tartu, Observatooriumi 1, 61602, Tõravere, Estonia}\\
$^{10}$Instituto de Astrof\'{i}sica e Ci\^{e}ncias do Espaço - Centro de Astrof\'isica da Universidade do Porto, Rua das Estrelas, 4150-762 Porto, Portugal\\
$^{11}${School of Mathematics and Physics, University of Queensland, St Lucia, Queensland 4072, Australia}\\
$^{12}${Institut d’Astrophysique de Paris, UMR 7095, CNRS, and Sorbonne Université, 98 bis boulevard Arago, 75014 Paris, France}\\
$^{13}$Universitäts-Sternwarte, Fakultät für Physik, Ludwig-Maximilians-Universität München, Scheinerstr. 1, 81679 München, Germany\\
$^{14}${Centre for Astrophysics \& Supercomputing, Swinburne University
of Technology, Hawthorn, VIC 3122, Australia}\\
$^{15}${University of Strasbourg, CNRS UMR 7550, Observatoire astronomique de Strasbourg, F-67000 Strasbourg, France}\\
$^{16}${SISSA -- International School for Advanced Studies, Via Bonomea 265, I-34136 Trieste, Italy}\\
$^{17}${UWC -- University of the Western Cape, Department of Physics and Astronomy, Cape Town 7535, South Africa}\\
$^{18}${IFPU -- Institute for Fundamental Physics of the Universe, Via Beirut, 2, 34151 Trieste, Italy}\\
$^{19}${INFN -- Sezione di Trieste, via Valerio 2, I-34127 Trieste, Italy}\\
$^{20}${Department of Astrophysics, University of Vienna, T\"urkenschanzstra{\ss}e 17, 1180 Vienna, Austria}\\
$^{21}${INAF -- Osservatorio Astronomico di Trieste, Via Tiepolo 11, I-34131 Trieste, Italy}\\

%%%%%%%%%%%%%%%%% APPENDICES %%%%%%%%%%%%%%%%%%%%%

\appendix

\section{Impact of cosmic web tracer selection on spin--filament alignments}
\label{Impact of cosmic web tracer selection on spin--filament alignments}
We test galaxy spin--filament alignments for the different tracer populations used to reconstruct the cosmic web and described in Section~\ref{Tracer populations: the combined redshift sample}, in order to assess whether the presence of a transition mass is robust to the adopted tracer selection. We compare the mass-matched constant-number-density approach, which is the fiducial tracer selection adopted in this work, with the full sample, the fixed $M_{\star}$ cut, and the weights approaches. The redshift-dependent $M_{\star}$ cut and the constant-number-density samples show larger residual variations in tracer-population consistency, poorer density uniformity, or larger tracer separations. As these selections perform poorly in reconstructing the cosmic web, they are excluded from the comparison.

\begin{figure}
\centering
%\vspace*{-18pt}
\includegraphics[scale=0.5]{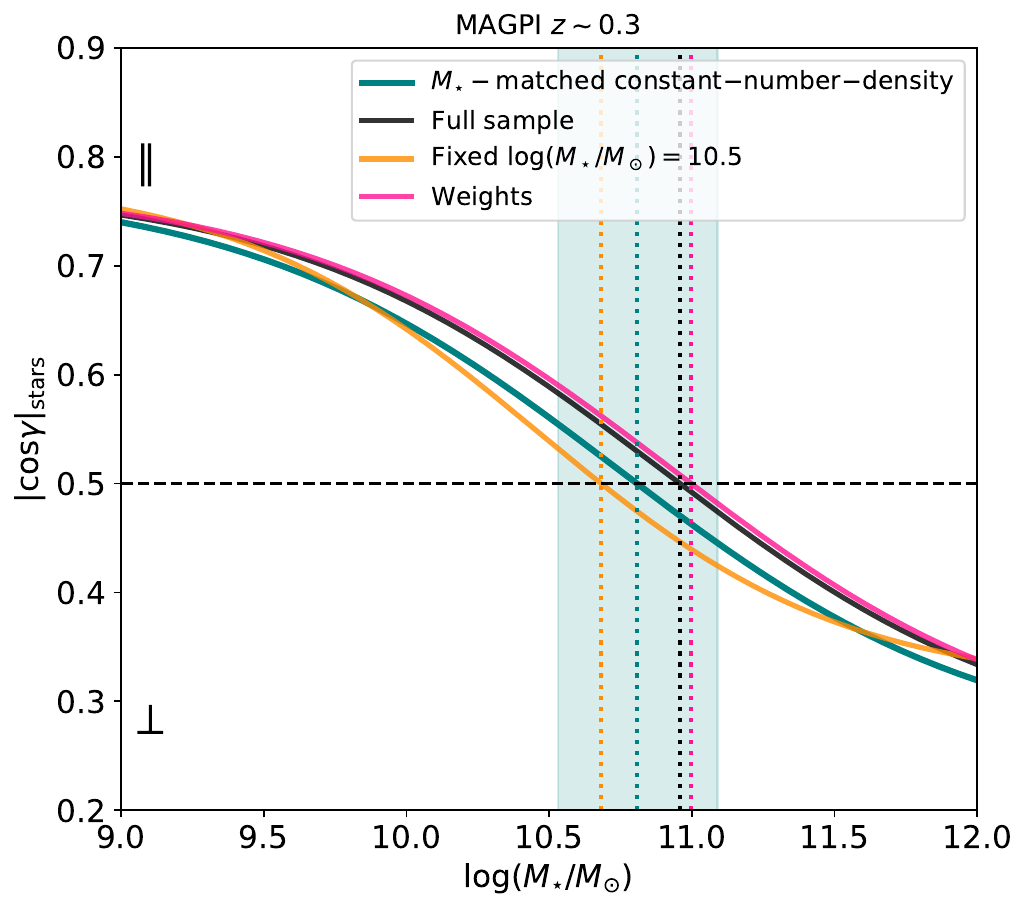}
\caption{|$\cos\gamma$| versus stellar mass for 342 MAGPI galaxies with the solid lines showing the fitted sigmoid functions for the mass-matched constant-number-density (teal), full sample (black), fixed $M_{\star}$ cut (orange) and weights (pink) tracer-selection approaches. The black dashed line marks $|\cos\gamma|=0.5$, the random orientation with respect to filaments. Dotted lines mark the corresponding transition stellar masses, with 1$\sigma$ uncertainty on the fiducial transition mass shown by the shaded teal region.}
\label{cosgamma_mass_CWtracers}
\end{figure}

The presence of a transition stellar mass remains robust across the tracer selections considered, with only modest insignificant shifts of $\sim0.13$--$0.19$ dex relative to the fiducial value at $<1\sigma$ level. Figure~\ref{cosgamma_mass_CWtracers} shows the fitted sigmoid relations for the different tracer-selection methods, together with their corresponding transition stellar masses, for the 342 MAGPI galaxies. The solid lines show the best-fitting sigmoid functions, while the teal dotted line indicates the transition stellar mass for the fiducial mass-matched constant-number-density sample, $\log(M_{\star,t}/M_{\odot})_{\rm MAGPI}=10.81\pm0.28$ (as reported in Figure~\ref{cosgamma_mass}). The shaded region denotes its $1\sigma$ uncertainty estimated as described in Section~\ref{Trends with stellar mass}. The transition masses obtained using the full sample, fixed $M_{\star}$ cut, and weights approaches are $\log(M_{\star,t}/M_{\odot})=10.96\pm0.18$, $10.68\pm0.30$, and $11.0\pm0.20$, corresponding to shifts of $+0.15$, $-0.13$, and $+0.19$ dex, respectively. Thus, although the exact location of the stellar mass transition depends on the tracer selection, the values are consistent within $1\sigma$, and the overall trend from preferentially parallel spin--filament alignments at low stellar masses to preferentially perpendicular alignments at high stellar masses is preserved.

%%%%%%%%%%%%%%%%%%%%%%%%%%%%%%%%%%%%%%%%%%%%%%%%%%

% Don't change these lines
\bsp	% typesetting comment
\label{lastpage}
\end{document}